\documentclass[onecolumn,english,aip,preprintnumbers,amsmath,amssymb,superscriptaddress]{revtex4-1}

\usepackage{graphicx}
\usepackage{dcolumn}
\usepackage{bm}
\usepackage[version=3]{mhchem}
\usepackage{color}
\usepackage{float}
\usepackage{bigstrut}

\begin{document}
\sloppy

\preprint{AIP/123-QED}

\title[Superconducting order parameter of the nodal-line semimetal NaAlSi]{Superconducting order parameter of the nodal-line semimetal NaAlSi}

\author{Lukas Muechler}
\affiliation{Center for Computational Quantum Physics, The Flatiron Institute, New York, New York 10010, USA}
 
\author{Zurab Guguchia}
\affiliation{Laboratory for Muon Spin Spectroscopy, Paul Scherrer Institute, CH-5232 Villigen PSI, Switzerland}%

\author{Jean-Christophe Orain}
\affiliation{Laboratory for Muon Spin Spectroscopy, Paul Scherrer Institute, CH-5232 Villigen PSI, Switzerland}%

\author{J\"urgen Nuss}
\affiliation{Max Planck Institute for Solid State Research, Heisenbergstr. 1, 70569 Stuttgart, Germany}%

\author{Leslie M. Schoop}
\affiliation{Department of Chemistry, Princeton University, 08544 Princeton, NJ, USA}%

\author{Ronny Thomale}
\affiliation{Institute for Theoretical Physics and Astrophysics, Julius-Maximilians University of W\"urzburg, 97074 W\"urzburg, Germany}

\author{Fabian O. von Rohr}
\affiliation{Department of Chemistry University of Zurich, CH-8057 Z\"urich, Switzerland }
\affiliation{Department of Physics, University of Zurich, CH-8057 Z\"urich, Switzerland}

%


\begin{abstract}
\textbf{Abstract.} Nodal-line semimetals are topologically non-trivial states of matter featuring band crossings along a closed curve, i.e. nodal-line, in momentum space. Through a detailed analysis of the electronic structure, we show for the first time that the normal state of the superconductor NaAlSi, with a critical temperature of $T_{\rm c} \approx$ 7 K, is a nodal-line semimetal, where the complex nodal-line structure is protected by non-symmorphic mirror crystal symmetries. We further report on muon spin rotation experiments revealing that the superconductivity in NaAlSi is truly of bulk nature, featuring a fully gapped Fermi-surface. The temperature-dependent magnetic penetration depth can be well described by a two-gap model consisting of two $s$-wave symmetric gaps with $\Delta_1 =$ 0.6(2) meV and $\Delta_2 =$ 1.39(1) meV. The zero-field muon experiment indicates that time-reversal symmetry is preserved in the superconducting state. Our observations suggest that notwithstanding its topologically non-trivial band structure, NaAlSi may be suitably interpreted as a conventional London superconductor, while more exotic superconducting gap symmetries cannot be excluded. The intertwining of topological electronic states and superconductivity renders NaAlSi a prototypical platform to search for unprecedented topological quantum phases.
\end{abstract}

\keywords{Nodal-line semimetals, topological semimetal, superconductor, $\mu$SR}
\maketitle

\section{Introduction}
In the course of the advent of topological insulators, several classes of non-trivial topological semimetals have been proposed and experimentally sought after: Weyl semimetals, Dirac semimetals, and nodal-line semimetals \cite{Weyl,Weyl_2,Dirac,nodal,nodal_2}. In their essence, all these types of topological matter arise from band inversion, often along with non-symmorphic symmetries. As opposed to insulators, the topological nature of a semimetal is given by a more intricate version of topological bulk invariants and bulk boundary correspondence. For the most elementary instance, the surface Fermi arcs of a Weyl semimetal are localized away from the Weyl cone projection points at the surface, and derive their localization length scale from the direct gap between the underlying bulk and valence bands at the given surface momentum. For nodal-line semimetals, the surface states feature intriguing structures referred to as drumhead states. Additional complexity can arise if the closed nodal-line takes on more complicated forms in momentum space, which yields structures called nodal knot semimetals. The higher the semimetallic topological complexity the harder it appears to find quantum matter realizations of such states, so classical metamaterial platforms often seem to be the only viable alternative\cite{Lee2019,Yang2019}.

In many respects, however, only quantum material realizations of topological semimetals promise a complete unfolding of their rich phenomenology\cite{Sur2016,Laubach2016}. The interplay of emergent quantum effects such as magnetism, charge-ordering, and superconductivity, together with nontrivial band topology has been identified as a promising platform for the realization of exotic quasi particles, such as e.g. Majorana fermions \cite{Majorana,Shapourian2018,Maj_2,Maj_3}. Recently, several candidates for topological materials with bulk superconductivity have been experimentally realized. Among those are topological insulators, namely \ce{Cu_{x}Bi2Se3}\cite{cuxBi2Se3}, or \ce{Tl5Te3}\cite{Tl5Te3}, Dirac semimetals, namely \ce{Cd3As2}\cite{Cd3As2},Au$_2$Pb~\cite{Schoop2015Dirac} and 2M-\ce{WS2}\cite{WS2_1,WS2_2}, Weyl semimetals such as \ce{MoTe2}\cite{MoTe2_1,MoTe2_2}, and also the nodal-line semimetal \ce{PbTaSe2}\cite{PbTaSe2}. Superconducting instabilities of nodal-line semimetals promise to be a particularly interesting class of compounds, as the nodal-lines can induce a Berry phase picked up for a closed path along the Fermi surface and thus constrain possible electronic pairing \cite{nodal_line_SC}. The only known nodal-line semimetal superconductor, i.e. \ce{PbTaSe2}\cite{PbTaSe2_2}, corresponds to an intercalated derivative of \ce{TaSe2}. This parent compound is a platform for complex charge ordering. Upon intercalation of lead the charge ordering is suppressed and superconductivity is observed at a critical temperature of $T_{\rm c} \approx$ 3.8 K\cite{PbTaSe2_3}.

The  ternary compound NaAlSi investigated here crystallizes in the centrosymmetric space group $P4/nmm$ of PbFCl structure-type, as shown in Fig~\ref{figband}(a). This compound is isostructural to the "111" Fe-based superconductors LiFeAs and LiFeP \cite{NaAlSi_structure}. NaAlSi has been reported to be a type-II superconductor with a critical temperature of $T_{\rm c} \approx$ 7 K at ambient pressure~\cite{NaAlSi_2}, while the isostructural and isoelectronic NaAlGe is not superconducting. NaAlSi and NaAlGe have almost exactly the same band structure except for one missing piece of small Fermi surface. This small Fermi surface is rather unusual, and further obscured by the small but seemingly important interlayer coupling along the crystalline c-axis\cite{NaAlSi_Pickett}. Even though one expects the phonon spectra for these two compounds to be fairly similar, and as such the onset of superconductivity if phonons were responsible for the pairing, the isoelectric cousin NaAlGe does not exhibit any kind of superconductivity down to low temperatures. 

H. B. Rhee \textit{et al.} have shown by first-principles  electronic  structure  calculations that  NaAlSi is what they refer to as "a naturally self-doped semimetal", with charge-transfer between the covalent bands within the substructure and two-dimensional free-electron-like bands within the Al-Si layers\cite{NaAlSi_Pickett}. This electronic structure results in an unusually small Fermi surface and a very low density of states at the Fermi level. Both characteristics, together with the reasonably high critical temperature of NaAlSi, are contradictory to conventional BCS theory predictions, where the critical temperature $T_{\rm c}$ depends expotentially on the density of states at the Fermi-level. Given these special electronic features, it was proposed that the superconductivity in this material may not be of phononic origin. Hence, the microscopic origin of superconductivity still remains to be unambiguously identified. It calls for a detailed study of the electronic structure of this nodal-line semimetal, and in particular of how small electronic deviations between NaAlGe and NaAlSi might affect unconventional pairing tendencies from a weak coupling perspective. An unconventional pressure dependence of the superconductivity in NaAlSi has been reported: The critical temperature was found to slightly increase up to a transition temperature of $T_{\rm c} \approx$ 9 K under an external pressure, and to disappear abruptly at a pressure of $p =$ 4.8 GPa in the absence of a structural transition~\cite{Schoop2012Effect}. 

Here, we report on detailed band structure calculations, showing that NaAlSi is a nodal-line semimetal for the first time, and on its superconducting order parameter analysis. Our muon spin rotation measurements reveal that the superconductivity in NaAlSi is truly of bulk nature and that the Fermi-surface is fully gapped with an $s+s$-wave symmetrical gap in the absence of time-reversal-symmetry breaking. Our observations suggest that superconductivity in this topologically non-trivial material may be explained as a conventional London superconductor. The results, however, do not exclude some more exotic superconducting gap symmetries either, which we will also briefly discuss in this work.

\section{Methods}

\subsection{DFT Calculations}

The DFT calculations were performed using the \textit{VASP} package~\cite{VASP} using the standard pseudopotentials for Na, Si and Al. The experimental geometries were taken from Ref.~\cite{Schoop2012Effect}. The reducible Brillouin zone was sampled by a $9\times9\times9$ k-mesh for the self-consistent calculations. A Wannier interpolation using 18 bands was performed by projecting onto an atomic-orbital basis centered at the atomic positions, consisting of Na-$3s$,Al-$3s$ and $3p$ as well as Si-$3s$ and $3p$ orbitals.
The nodal-lines were calculated via the package \textit{wanniertools}~\cite{wu2018wanniertools} based on this Wannier interpolation.

\subsection{Sample Preparation}

Blue-metallic, highly crystalline samples of NaAlSi were prepared by solid-state synthesis. In a first step, the elements Na (purity 99.99 \%), Al (purity 99.999 \%) and Si (purity 99.9999 \%) were mixed in a ratio of Na:Al:Si = 3:1:1. The excess of Na is necessary for obtaining phase pure products, it partly acts as a sodium flux. This mixture (2g) was sealed in an argon-filled tantalum tube in order to minimize Na loss during the reaction. The tantalum tube was sealed in a quartz ampule in order to prevent the oxidation of the tantalum reaction vessel. The reaction was carried out at 700 $^\circ$C for three days (heating rate 50 $^\circ$C/h). To increase the crystallinity of the product, the sample was slowly cooled (5 $^\circ$C/h) to 600 $^\circ$C, held for three days at this temperature, and finally cooled to room temperature (5 $^\circ$C/h). In a second step, the excess sodium was removed by distillation of the product under a dynamic vacuum at 200 $^\circ$C for 100 h. The phase purity and crystal structure of the  sample  was  verified  by  powder  and  single  crystal  x-ray diffraction  using  a  D8  Focus  diffractometer  with  Cu K$_{\alpha 1}$ radiation (Bruker AXS GmbH, Karlsruhe, Germany).
 
\subsection{$\mu$SR Measurements}

Spin-polarized  muons ($\mu^{+}$) are extremely sensitive local magnetic probes that were here used to investigate the field distribution of the vortex state in the type-II superconductor NaAlSi. Transverse field (TF) and zero field (ZF) $\mu$SR experiments were carried down to $T =$ 250 mK, well below the superconducting transition temperature of NaAlSi. Pressed pellets of NaAlSi were transferred under inert atmosphere with a portable glovebox into the cryostat. The $\mu$SR spectra have been analyzed using the MUSRFIT software package\cite{MUSR}.

\section{Results and Discussion}

\subsection{Electronic Structure of NaAlSi}
\begin{figure}
\centering
\includegraphics[width=0.98\linewidth]{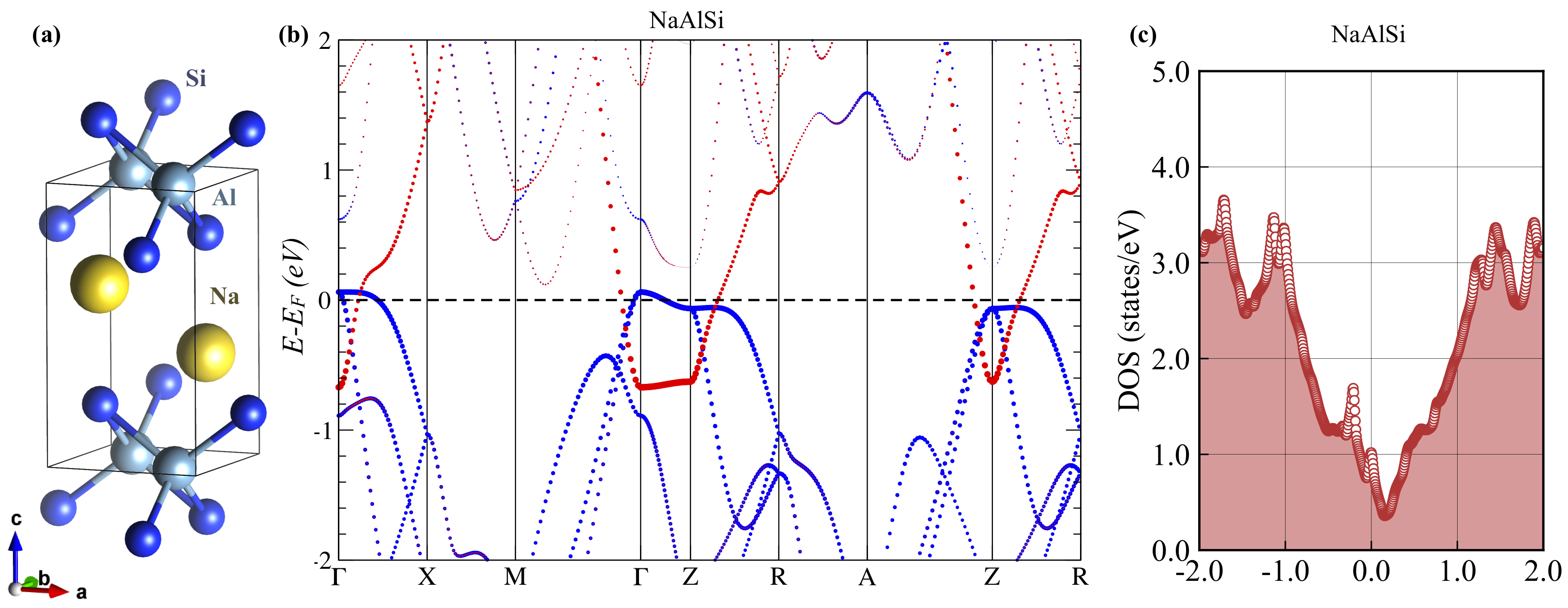}
\caption{
(a) Crystal structure of NaAlSi (b) Band structure of NaAlSi along high-symmetry points. Blue color indicates contributions from Si-$p$-orbitals whereas red color indicates Al-$p$-character. The size of the dots is proportional to the contribution of the orbitals at each k-point. (c) Density of states of bulk NaAlSi between -2 eV to 2 eV from the Fermi-level.} 
\label{figband}
\end{figure}

Charge balanced materials with 8 valence electrons of the form $X$AlSi ($X=$Li, Na) are expected to be semiconductors due to a completely filled Si $p-$shell that is separated by a gap from the empty $s$-shells of Li$^{+}$ and Al$^{3+}$.
While LiAlSi, which crystallizes in the cubic Half-Heusler structure, is indeed a semiconductor with a large direct band gap, NaAlSi shows metallic transport behaviour and becomes superconducting below 7 K. This can be attributed to the fact that NaAlSi does not crystallize in a Half-Heusler structure, but in the tetragonal spacegroup $P4/nmm$, which consists of edge-sharing AlSi$_4$ tetrahedra with short Al-Al distances of about 2.9 \AA. 

The short distance leads to considerable bonding between the Al-atoms in the $x-y$ plane, which is indicated by a dispersive Al-$s$-band close to the Fermi energy ($E_F$) with a bandwidth of over 4 eV as shown in Fig~\ref{figband}(b). As expected from the ionic picture discussed above, the band structure close to the Fermi level consists mostly of bands derived from the Si-$p$-orbitals, which are almost completely filled. However, in contrast to LiAlSi, the large bandwidth with of the Al-$s$-bands leads to a band inversion between the two sets of bands and results in a metallic band structure with multiple linear band crossings close to $E_F$ in NaAlSi. These band crossings form a complex nodal-line structure, protected by the crystalline symmetries of the space group $P4/nmm$, namely the non-symmorphic mirror symmetry $\bar{M}_z = \{ M_z | \frac{1}{2}\frac{1}{2}0\}$, the mirror $M_{xy}$, spatial inversion $\{I | 0 0 0\}$ and the two screw symmetries $\bar{C}_{2x} = \{C_{2x} | \frac{1}{2} 0 0 \}$ and $\bar{C}_{2y} = \{C_{2y} |0 \frac{1}{2} 0 \}$.
While nodal-lines and band inversion are a common feature in this space group, the origin of the nodal-lines in NaAlSi is of different origin than those found in e.g. the ZrSiS-family~\cite{schoop2016dirac,pezzini2018unconventional} and FeTe$_{1-x}$Se$_x$~\cite{zhang2018observation}.

\begin{figure}
\centering
\includegraphics[width=0.8\linewidth]{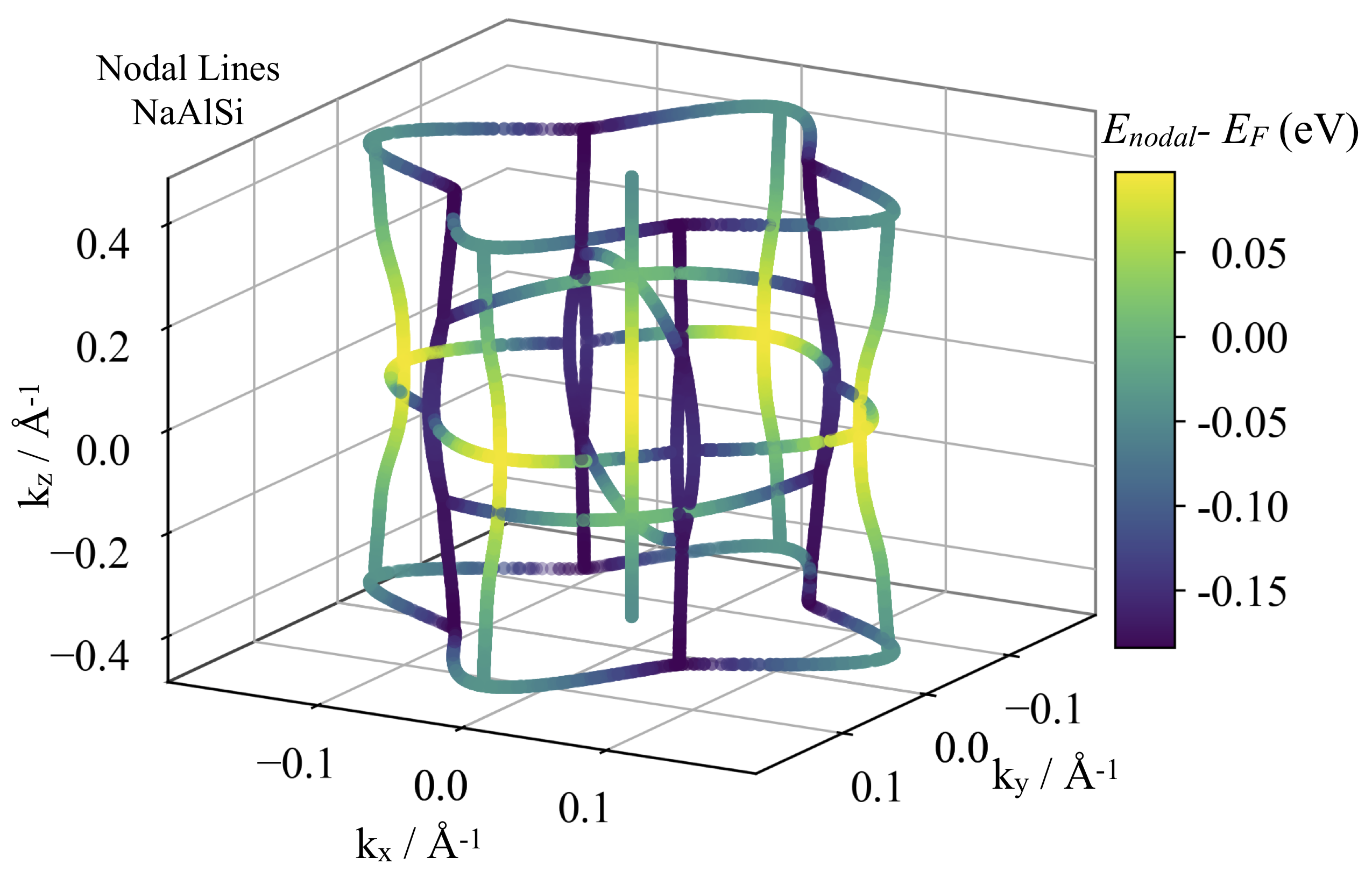}
\caption{Nodal-lines derived from the overlap of the highest energy Si-$p$-band with the Al-$s$-band. The energy of the crossing point is given relative to $E_F$.} 
\label{fig_nodal}
\end{figure}

Figure~\ref{fig_nodal} shows the resulting nodal-line structure. It consists of two nodal-lines in the $k_z = 0$ and the $k_z = \pi$ plane, which are interconnected by nodal-lines along the $k_z$ direction. The nodal-line shows dispersion, i.e. the gap closing points do not all occur at one energy, but in a window of about 200 $meV$ around $E_F$.

\subsection{Temperature dependent magnetic penetration depth}

In figure \ref{fig2}(a), we show the transverse-field (TF) ${\mu}$SR-time spectra for NaAlSi, measured in an applied magnetic field of ${\mu}_{\rm 0}H = 20$~mT. The spectra above (8 K) and below (0.25 K) the superconducting transition temperature $T_{{\rm c}}$ are depicted. In the normal state, the oscillations show a small relaxation due to the random local fields from the nuclear magnetic moments. Below the critical temperature $T_{{\rm c}}$ the relaxation rate strongly increases with decreasing temperature due to the presence of a nonuniform local magnetic field distribution as a result of the formation of a flux-line lattice (FLL) in the superconducting state. This is the first reported evidence, in the absence of specific heat measurements, for the bulk nature of the superconductivity in this material.

Magnetism, if present in the samples, may also enhance the muon depolarization rate. Therefore, we have carried out ZF-${\mu}$SR experiments above and below $T_{{\rm c}}$ to search for magnetism (static or fluctuating) in NaAlSi. As shown in figure \ref{fig2}b no sign of either static or fluctuating magnetism could be detected in ZF time spectra down to 0.25 K. The spectra are well described by a weakly damped Kubo-Toyabe depolarization function \cite{Toyabe}, reflecting the field distribution at the muon site created by the nuclear and weak electronic moments. Moreover, no change in ZF-${\mu}$SR relaxation rate (see figure \ref{fig2}c) across $T_{c}$ was observed, pointing to the absence of any spontaneous magnetic fields associated with time-reversal symmetry breaking pairing state in NaAlSi (compare references \onlinecite{LukeTRS,HillierTRS,BiswasTRS}).

The temperature dependence of the muon spin depolarization rate ${\sigma}_{{\rm sc}}$, which is proportional to the second moment of the field distribution (see Method section), of NaAlSi in the superconducting state is shown in figure \ref{fig2}d. Below $T_{{\rm c}}$ the relaxation rate ${\sigma}_{{\rm sc}}$ starts to increase from zero with decreasing temperature due to the formation of the FLL. Interestingly, the form of the temperature dependence of ${\sigma}_{{\rm sc}}$, which reflects the topology of the superconducting gap, shows the saturation upon lowering the temperature below approximately 2 K. We show in the following how these behaviours indicate the presence of the two isotropic $s$-wave gaps on the Fermi surface of NaAlSi.

\subsection{Probing the nonuniform field distribution in the vortex state} 

\begin{figure*}
\centering
\includegraphics[width=0.9\linewidth]{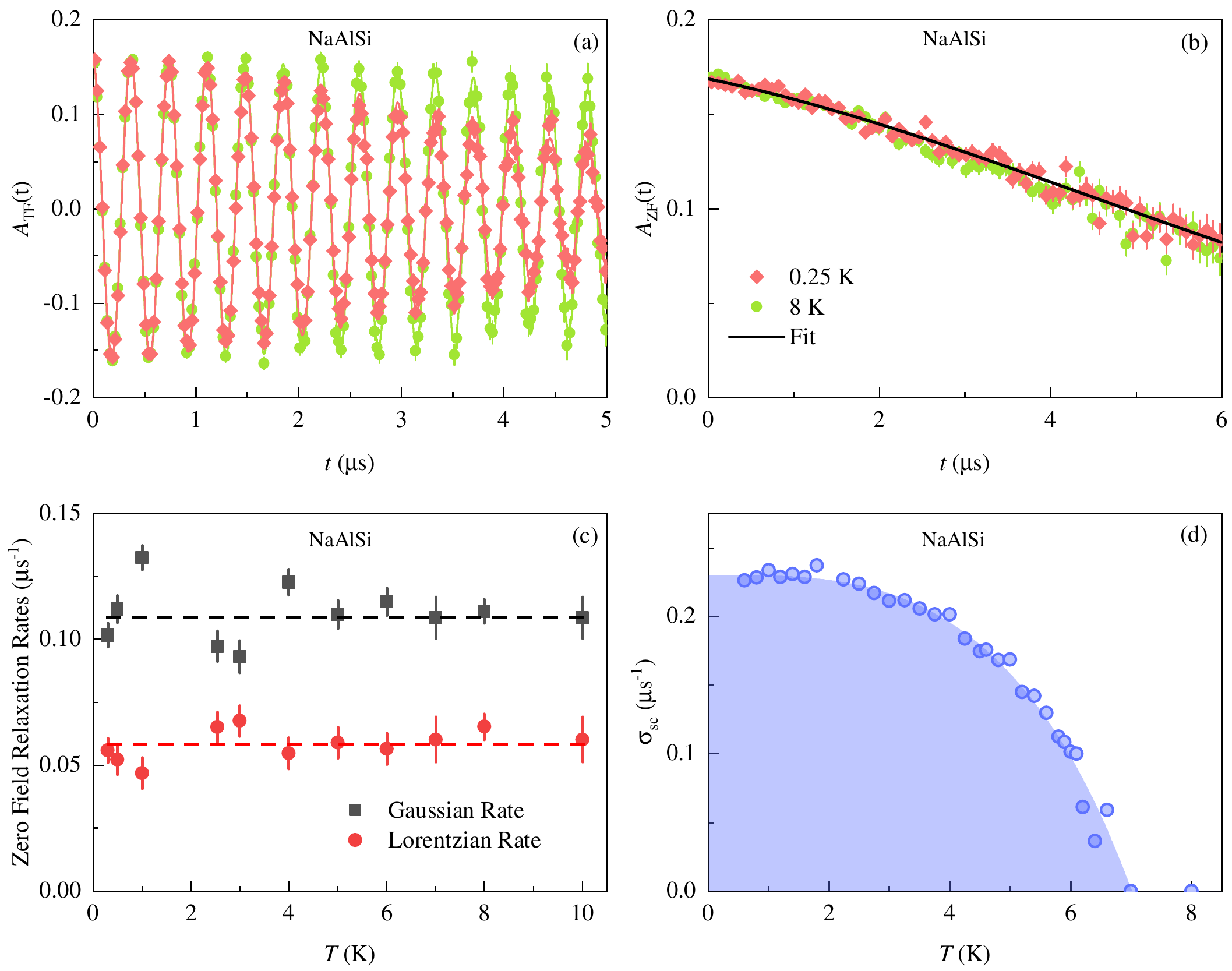}
\caption{(a) The transverse-field ${\mu}$SR time spectra for NaAlSi, obtained above and below $T_{\rm c}$ (after field cooling the sample from above $T_{\rm c}$). The solid lines in represent fits to the data. (b) Zero-field ${\mu}$SR time spectra for NaAlSi recorded above and below $T_{\rm c}$. The line represents the fit to the data of the combination of Lorentzian and Gaussian Kubo-Toyabe depolarization function \cite{Toyabe}. (c) The temperature dependence of the ZF Gaussian and Lorentzian depolarization rates. (d) The temperature dependence of the superconducting muon spin depolarization rate ${\sigma}_{\rm sc}$ for NaAlSi, measured in an applied magnetic field of ${\mu}_{\rm 0}H = 20$~mT.} 
\label{fig2}
\end{figure*}

\begin{figure}
\centering
\includegraphics[width=0.5\linewidth]{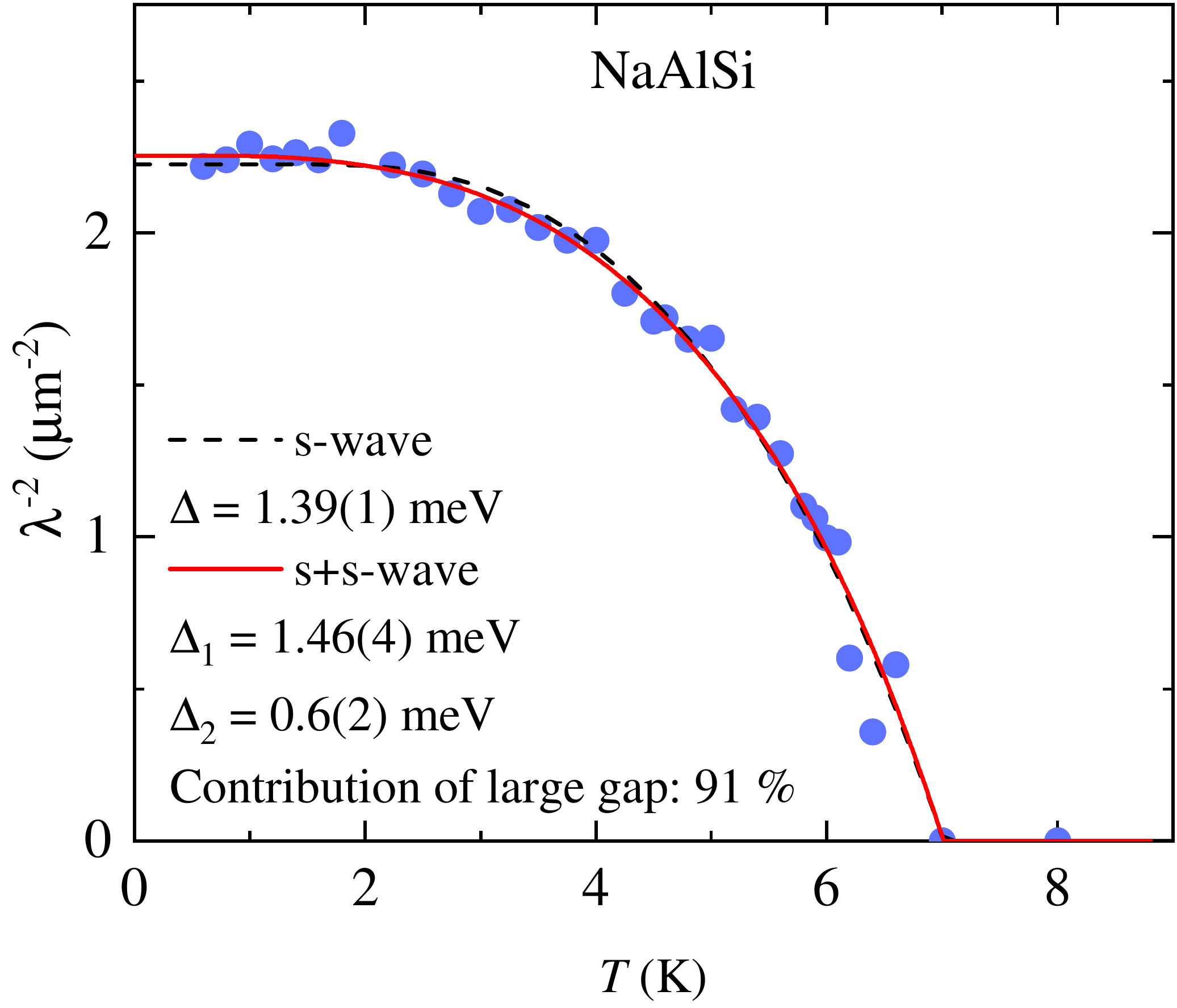}
\caption{The temperature dependence of ${\lambda}^{-2}$ for NaAlSi. The solid line corresponds to a two-gap (s+s)-wave model and the dashed line represents a fit using a single gap $s$-wave model.} 
\label{fig3}
\end{figure}

In order to investigate the symmetry of the superconducting gap, we note that ${\lambda}(T)$ is related to the relaxation rate ${\sigma}_{{\rm sc}}(T)$ by the equation \cite{Brandt} 
\begin{equation}
\frac{\sigma_{sc}(T)}{\gamma_{\mu}}=0.06091\frac{\Phi_{0}}{\lambda^{2}(T)},
\end{equation}

where ${\gamma_{\mu}}$ is the gyromagnetic ratio of the muon, and ${\Phi}_{{\rm 0}}$ is the magnetic-flux quantum. Thus, the flat $T$-dependence of ${\sigma}_{{\rm sc}}$ observed for low temperatures (see figure \ref{fig2}(d)) is consistent with a nodeless, fully-gapped superconductor, in which $\lambda^{-2}\left(T\right)$ reaches its zero-temperature value exponentially.

To proceed with a quantitative analysis, we consider the local (London) approximation (${\lambda}$ ${\gg}$ ${\xi}$, where ${\xi}$ is the coherence length) and employ the empirical ${\alpha}$-model. The model, widely used in previous investigations of the penetration depth of multi-band superconductors \cite{Tinkham,carrington,padamsee,MgB2,NbSe2} assumes that the gaps occurring in different bands, besides a common
$T_{{\rm c}}$, are independent of each other. The superfluid density is calculated for each component separately \cite{GuguchiaNature} and added together with a weighting factor. For our purposes, a two-band model suffices, yielding

\begin{equation}
\frac{\lambda^{-2}(T)}{\lambda^{-2}(0)}=\omega_{1}\frac{\lambda^{-2}(T,\Delta_{0,1})}{\lambda^{-2}(0,\Delta_{0,1})}+\omega_{2}\frac{\lambda^{-2}(T,\Delta_{0,2})}{\lambda^{-2}(0,\Delta_{0,2})},
\end{equation}

where ${\lambda}(0)$ is the penetration depth at zero temperature, ${\Delta_{0,i}}$ is the value of the $i$-th superconducting gap ($i=1$, 2) at $T=0$~K, and ${\omega}_{i}$ is the weighting factor which measures their relative contributions to ${\lambda^{-2}}$ (i.e. ${\omega}_{1}+{\omega}_{2}=1$).

The results of this analysis are presented in figure \ref{fig3}, where the temperature dependence of ${\lambda^{-2}}$ for NaAlSi is plotted. We consider two different possibilities for the gap functions: either a constant gap, $\Delta_{0,i}=\Delta_{i}$, or an angle-dependent gap of the form $\Delta_{0,i}=\Delta_{i}\cos2\varphi$, where $\varphi$ is the polar angle around the Fermi surface. The dashed and the solid lines represent a fit to the data using an $s$-wave and an $s+s$-wave model, respectively. The two gap $s$ + $s$-wave scenario with a small gap ${\Delta}_{1}$ ${\simeq}$ 0.6(2) meV  and a large gap ${\Delta}_{2}$ ${\simeq}$ 1.39(1) meV (with the weighting factor of ${\omega}_{2}$ = 0.91), describes the experimental data slightly better than the single gap $s$-wave model, while the relative weight of the small gap is only approximately 10 \%. This result might also correspond to a single anisotropic gap with $s$-wave symmetry, this analysis is insensitive to these different scenarios (compare, e.g., references \onlinecite{padamsee,MoTe2_2,MgB2,NbSe2}). 

Even though a phonon-driven mechanism for superconductivity appears likely, and from there a conventional $s$-wave type order parameter, the experimental evidence reported by us for NaAlSi raises the question of what kind of superconducting order parameter symmetries can in principle be imagined for such a compound. The reported bulk gap excludes nodes in the superconducting phase, and the absence of a zero field signal in $\mu$SR strongly implies preserved time reversal symmetry in the superconductor. The given space group, combined with the co-dimension $p=2$ of the Fermi surface in the normal state, does not exclude alternative pairings to trivial $s$-wave.
Naively, the point group $D_{4h}$ of NaAlSi possesses three irreducible representations, $A_{1g}$, $B_{1g}$ and $B_{2g}$, that are consistent with singlet paring in the case of vanishing spin-orbit coupling. Since all evidence disfavors line nodes in the superconducting phase and time-reversal symmetry is kept, $A_{1g}$ is the likeliest of a possible two-dimensional irreducible representations. In terms of unconventional pairing, the options are therefore reduced to very excotic superconducting symmetries, such as e.g. helical superconductors, or (slightly more realistically) a relative sign change between the two $s$-wave sheets. The experiments at hand cannot alone distinguish between sign changing s$^{+-}$, hence topological, and s$^{++}$ trivial pairing states. Generally, s$^{++}$ can be considered as more likely. However, the high sensitivity of the superconducting state in NaAlSi to disorder, might suggest that a s$^{+-}$ might be realized in this material. Further surface probes such as, e.g., the Kerr effect or phase sensitive measurements, however, would be needed to reach conclusive statements on such questions. \\

\section{Conclusion}

In summary, we have reported on a detailed analysis of the electronic structure of the compound NaAlSi. We here describe for the first time that the superconductor NaAlSi has a topological non-trivial nodal-line band structure. Its complex nodal-line structure is thereby protected by the symmetry of its crystal structure, in particular by the non-symmorphic mirror symmetries of space group $P4/nmm$. We have characterized the microscopic superconducting properties of NaAlSi by a series of $\mu$SR experiments. The TF $\mu$SR experiments reveal for the first time unambiguously that the superconductivity in NaAlSi is of bulk nature. The measured temperature-dependant magnetic penetration depth $\lambda$ corresponds to a fully-gapped Fermi-surface. In our analysis an $s+s$-wave symmetrical gap with $\Delta_1 =$ 0.6(2) meV and $\Delta_2 =$ 1.39(1) meV was sufficient to explain the observed behavior. The ZF $\mu$SR experiments above and below the critical temperature indicated the preservation of time reversal symmetry. 

Our results indicate that superconductivity in this topologically non-trivial material may be explained as a conventional London superconductor. These results, however, also do not exclude some more exotic superconducting gap symmetries, such as e.g. a s$^{+-}$ symmetric gap. It may be speculated, whether the observed disappearance of superconductivity under pressure or the absence of superconductivity in isoelectronic and isostructural NaAlGe could be tied to the change in topological class in this material. Further theoretical as well as experimental (especially measurements of the bandstructure and thermodynamic measurements of the superconducting properties) work are crucial for understanding the interplay of superconductivity and topology in this material. We expect the results at hand to generally motivate significant additional studies into materials that couple topology to emergent quantum effects.

\section{Acknowledgement}

We thank Sabine Prill-Diemer (Max Planck Institute, Stuttgart) for help with the synthesis, Yan Sun for help related to the creation of the Wannier functions, as well as Mark H. Fischer for helpful discussion. The ${\mu}$SR experiments were carried out at the Swiss Muon Source (S${\mu}$S) Paul Scherrer Insitute, Villigen, Switzerland. This  work  was  supported  by  the  Swiss National Science Foundation under Grant No. PZ00P2\_174015 and the Ernst G\"ohner Fellowship 2019 awarded by the "Fond zur F\"orderung des akademischen Nachwuchs" (FAN). The Flatiron Institute is a division of the Simons Foundation. The work in W\"urzburg is funded by the Deutsche Forschungsgemeinschaft (DFG, German Research Foundation) through Project-ID 258499086 - SFB 1170 and through the W\"urzburg-Dresden Cluster of Excellence on Complexity and Topology in Quantum Matter -- \textit{ct.qmat} Project-ID 39085490 - EXC 2147. Work at Princeton was supported by NSF through the Princeton Center for Complex Materials, a Materials Research Science and Engineering Center DMR-1420541.
\textbf{Note:} This manuscript was submitted to \textit{APL Materials} on August 12, 2019.

\bibliography{lit}

\begin{thebibliography}{46}%
\makeatletter
\providecommand \@ifxundefined [1]{%
 \@ifx{#1\undefined}
}%
\providecommand \@ifnum [1]{%
 \ifnum #1\expandafter \@firstoftwo
 \else \expandafter \@secondoftwo
 \fi
}%
\providecommand \@ifx [1]{%
 \ifx #1\expandafter \@firstoftwo
 \else \expandafter \@secondoftwo
 \fi
}%
\providecommand \natexlab [1]{#1}%
\providecommand \enquote  [1]{``#1''}%
\providecommand \bibnamefont  [1]{#1}%
\providecommand \bibfnamefont [1]{#1}%
\providecommand \citenamefont [1]{#1}%
\providecommand \href@noop [0]{\@secondoftwo}%
\providecommand \href [0]{\begingroup \@sanitize@url \@href}%
\providecommand \@href[1]{\@@startlink{#1}\@@href}%
\providecommand \@@href[1]{\endgroup#1\@@endlink}%
\providecommand \@sanitize@url [0]{\catcode `\\12\catcode `\$12\catcode
  `\&12\catcode `\#12\catcode `\^12\catcode `\_12\catcode `\%12\relax}%
\providecommand \@@startlink[1]{}%
\providecommand \@@endlink[0]{}%
\providecommand \url  [0]{\begingroup\@sanitize@url \@url }%
\providecommand \@url [1]{\endgroup\@href {#1}{\urlprefix }}%
\providecommand \urlprefix  [0]{URL }%
\providecommand \Eprint [0]{\href }%
\providecommand \doibase [0]{http://dx.doi.org/}%
\providecommand \selectlanguage [0]{\@gobble}%
\providecommand \bibinfo  [0]{\@secondoftwo}%
\providecommand \bibfield  [0]{\@secondoftwo}%
\providecommand \translation [1]{[#1]}%
\providecommand \BibitemOpen [0]{}%
\providecommand \bibitemStop [0]{}%
\providecommand \bibitemNoStop [0]{.\EOS\space}%
\providecommand \EOS [0]{\spacefactor3000\relax}%
\providecommand \BibitemShut  [1]{\csname bibitem#1\endcsname}%
\let\auto@bib@innerbib\@empty
\bibitem [{\citenamefont {Weng}\ \emph {et~al.}(2015)\citenamefont {Weng},
  \citenamefont {Fang}, \citenamefont {Fang}, \citenamefont {Bernevig},\ and\
  \citenamefont {Dai}}]{Weyl}%
  \BibitemOpen
  \bibfield  {author} {\bibinfo {author} {\bibfnamefont {H.}~\bibnamefont
  {Weng}}, \bibinfo {author} {\bibfnamefont {C.}~\bibnamefont {Fang}}, \bibinfo
  {author} {\bibfnamefont {Z.}~\bibnamefont {Fang}}, \bibinfo {author}
  {\bibfnamefont {B.~A.}\ \bibnamefont {Bernevig}}, \ and\ \bibinfo {author}
  {\bibfnamefont {X.}~\bibnamefont {Dai}},\ }\bibfield  {title} {\enquote
  {\bibinfo {title} {{Weyl Semimetal Phase in Noncentrosymmetric
  Transition-Metal Monophosphides}},}\ }\href {\doibase
  10.1103/PhysRevX.5.011029} {\bibfield  {journal} {\bibinfo  {journal} {Phys.
  Rev. X}\ }\textbf {\bibinfo {volume} {5}},\ \bibinfo {pages} {011029}
  (\bibinfo {year} {2015})}\BibitemShut {NoStop}%
\bibitem [{\citenamefont {Xu}\ \emph {et~al.}(2015)\citenamefont {Xu},
  \citenamefont {Belopolski}, \citenamefont {Alidoust}, \citenamefont
  {Neupane}, \citenamefont {Bian}, \citenamefont {Zhang}, \citenamefont
  {Sankar}, \citenamefont {Chang}, \citenamefont {Yuan}, \citenamefont {Lee}
  \emph {et~al.}}]{Weyl_2}%
  \BibitemOpen
  \bibfield  {author} {\bibinfo {author} {\bibfnamefont {S.-Y.}\ \bibnamefont
  {Xu}}, \bibinfo {author} {\bibfnamefont {I.}~\bibnamefont {Belopolski}},
  \bibinfo {author} {\bibfnamefont {N.}~\bibnamefont {Alidoust}}, \bibinfo
  {author} {\bibfnamefont {M.}~\bibnamefont {Neupane}}, \bibinfo {author}
  {\bibfnamefont {G.}~\bibnamefont {Bian}}, \bibinfo {author} {\bibfnamefont
  {C.}~\bibnamefont {Zhang}}, \bibinfo {author} {\bibfnamefont
  {R.}~\bibnamefont {Sankar}}, \bibinfo {author} {\bibfnamefont
  {G.}~\bibnamefont {Chang}}, \bibinfo {author} {\bibfnamefont
  {Z.}~\bibnamefont {Yuan}}, \bibinfo {author} {\bibfnamefont {C.-C.}\
  \bibnamefont {Lee}},  \emph {et~al.},\ }\bibfield  {title} {\enquote
  {\bibinfo {title} {{Discovery of a Weyl fermion semimetal and topological
  Fermi arcs}},}\ }\href@noop {} {\bibfield  {journal} {\bibinfo  {journal}
  {Science}\ }\textbf {\bibinfo {volume} {349}},\ \bibinfo {pages} {613--617}
  (\bibinfo {year} {2015})}\BibitemShut {NoStop}%
\bibitem [{\citenamefont {Bradlyn}\ \emph {et~al.}(2016)\citenamefont
  {Bradlyn}, \citenamefont {Cano}, \citenamefont {Wang}, \citenamefont
  {Vergniory}, \citenamefont {Felser}, \citenamefont {Cava},\ and\
  \citenamefont {Bernevig}}]{Dirac}%
  \BibitemOpen
  \bibfield  {author} {\bibinfo {author} {\bibfnamefont {B.}~\bibnamefont
  {Bradlyn}}, \bibinfo {author} {\bibfnamefont {J.}~\bibnamefont {Cano}},
  \bibinfo {author} {\bibfnamefont {Z.}~\bibnamefont {Wang}}, \bibinfo {author}
  {\bibfnamefont {M.~G.}\ \bibnamefont {Vergniory}}, \bibinfo {author}
  {\bibfnamefont {C.}~\bibnamefont {Felser}}, \bibinfo {author} {\bibfnamefont
  {R.~J.}\ \bibnamefont {Cava}}, \ and\ \bibinfo {author} {\bibfnamefont
  {B.~A.}\ \bibnamefont {Bernevig}},\ }\bibfield  {title} {\enquote {\bibinfo
  {title} {Beyond dirac and weyl fermions: Unconventional quasiparticles in
  conventional crystals},}\ }\href@noop {} {\bibfield  {journal} {\bibinfo
  {journal} {Science}\ }\textbf {\bibinfo {volume} {353}} (\bibinfo {year}
  {2016})}\BibitemShut {NoStop}%
\bibitem [{\citenamefont {Burkov}, \citenamefont {Hook},\ and\ \citenamefont
  {Balents}(2011)}]{nodal}%
  \BibitemOpen
  \bibfield  {author} {\bibinfo {author} {\bibfnamefont {A.~A.}\ \bibnamefont
  {Burkov}}, \bibinfo {author} {\bibfnamefont {M.~D.}\ \bibnamefont {Hook}}, \
  and\ \bibinfo {author} {\bibfnamefont {L.}~\bibnamefont {Balents}},\
  }\bibfield  {title} {\enquote {\bibinfo {title} {Topological nodal
  semimetals},}\ }\href {\doibase 10.1103/PhysRevB.84.235126} {\bibfield
  {journal} {\bibinfo  {journal} {Phys. Rev. B}\ }\textbf {\bibinfo {volume}
  {84}},\ \bibinfo {pages} {235126} (\bibinfo {year} {2011})}\BibitemShut
  {NoStop}%
\bibitem [{\citenamefont {Bzdu{\v{s}}ek}\ \emph {et~al.}(2016)\citenamefont
  {Bzdu{\v{s}}ek}, \citenamefont {Wu}, \citenamefont {R{\"u}egg}, \citenamefont
  {Sigrist},\ and\ \citenamefont {Soluyanov}}]{nodal_2}%
  \BibitemOpen
  \bibfield  {author} {\bibinfo {author} {\bibfnamefont {T.}~\bibnamefont
  {Bzdu{\v{s}}ek}}, \bibinfo {author} {\bibfnamefont {Q.}~\bibnamefont {Wu}},
  \bibinfo {author} {\bibfnamefont {A.}~\bibnamefont {R{\"u}egg}}, \bibinfo
  {author} {\bibfnamefont {M.}~\bibnamefont {Sigrist}}, \ and\ \bibinfo
  {author} {\bibfnamefont {A.~A.}\ \bibnamefont {Soluyanov}},\ }\bibfield
  {title} {\enquote {\bibinfo {title} {Nodal-chain metals},}\ }\href@noop {}
  {\bibfield  {journal} {\bibinfo  {journal} {Nature}\ }\textbf {\bibinfo
  {volume} {538}},\ \bibinfo {pages} {75} (\bibinfo {year} {2016})}\BibitemShut
  {NoStop}%
\bibitem [{\citenamefont {Lee}\ \emph {et~al.}(2019)\citenamefont {Lee},
  \citenamefont {Hofmann}, \citenamefont {Helbig}, \citenamefont {Liu},
  \citenamefont {Zhang}, \citenamefont {Greiter},\ and\ \citenamefont
  {Thomale}}]{Lee2019}%
  \BibitemOpen
  \bibfield  {author} {\bibinfo {author} {\bibfnamefont {C.-H.}\ \bibnamefont
  {Lee}}, \bibinfo {author} {\bibfnamefont {T.}~\bibnamefont {Hofmann}},
  \bibinfo {author} {\bibfnamefont {T.}~\bibnamefont {Helbig}}, \bibinfo
  {author} {\bibfnamefont {Y.}~\bibnamefont {Liu}}, \bibinfo {author}
  {\bibfnamefont {X.}~\bibnamefont {Zhang}}, \bibinfo {author} {\bibfnamefont
  {M.}~\bibnamefont {Greiter}}, \ and\ \bibinfo {author} {\bibfnamefont
  {R.}~\bibnamefont {Thomale}},\ }\bibfield  {title} {\enquote {\bibinfo
  {title} {Imaging nodal knots in momentum space through topolectrical
  circuits},}\ }\href@noop {} {\bibfield  {journal} {\bibinfo  {journal} {arXiv
  preprint arXiv:1904.10183}\ } (\bibinfo {year} {2019})}\BibitemShut {NoStop}%
\bibitem [{\citenamefont {Yang}\ \emph {et~al.}(2019)\citenamefont {Yang},
  \citenamefont {Chiu}, \citenamefont {Fang},\ and\ \citenamefont
  {Hu}}]{Yang2019}%
  \BibitemOpen
  \bibfield  {author} {\bibinfo {author} {\bibfnamefont {Z.}~\bibnamefont
  {Yang}}, \bibinfo {author} {\bibfnamefont {C.-K.}\ \bibnamefont {Chiu}},
  \bibinfo {author} {\bibfnamefont {C.}~\bibnamefont {Fang}}, \ and\ \bibinfo
  {author} {\bibfnamefont {J.}~\bibnamefont {Hu}},\ }\bibfield  {title}
  {\enquote {\bibinfo {title} {Evolution of nodal lines and knot transitions in
  topological semimetals},}\ }\href@noop {} {\bibfield  {journal} {\bibinfo
  {journal} {arXiv preprint arXiv:1905.00210}\ } (\bibinfo {year}
  {2019})}\BibitemShut {NoStop}%
\bibitem [{\citenamefont {Sur}\ and\ \citenamefont
  {Nandkishore}(2016)}]{Sur2016}%
  \BibitemOpen
  \bibfield  {author} {\bibinfo {author} {\bibfnamefont {S.}~\bibnamefont
  {Sur}}\ and\ \bibinfo {author} {\bibfnamefont {R.}~\bibnamefont
  {Nandkishore}},\ }\bibfield  {title} {\enquote {\bibinfo {title}
  {{Instabilities of Weyl loop semimetals}},}\ }\href {\doibase
  10.1088/1367-2630/18/11/115006} {\bibfield  {journal} {\bibinfo  {journal}
  {New Journal of Physics}\ }\textbf {\bibinfo {volume} {18}},\ \bibinfo
  {pages} {115006} (\bibinfo {year} {2016})}\BibitemShut {NoStop}%
\bibitem [{\citenamefont {Laubach}\ \emph {et~al.}(2016)\citenamefont
  {Laubach}, \citenamefont {Platt}, \citenamefont {Thomale}, \citenamefont
  {Neupert},\ and\ \citenamefont {Rachel}}]{Laubach2016}%
  \BibitemOpen
  \bibfield  {author} {\bibinfo {author} {\bibfnamefont {M.}~\bibnamefont
  {Laubach}}, \bibinfo {author} {\bibfnamefont {C.}~\bibnamefont {Platt}},
  \bibinfo {author} {\bibfnamefont {R.}~\bibnamefont {Thomale}}, \bibinfo
  {author} {\bibfnamefont {T.}~\bibnamefont {Neupert}}, \ and\ \bibinfo
  {author} {\bibfnamefont {S.}~\bibnamefont {Rachel}},\ }\bibfield  {title}
  {\enquote {\bibinfo {title} {{Density wave instabilities and surface state
  evolution in interacting Weyl semimetals}},}\ }\href {\doibase
  10.1103/PhysRevB.94.241102} {\bibfield  {journal} {\bibinfo  {journal} {Phys.
  Rev. B}\ }\textbf {\bibinfo {volume} {94}},\ \bibinfo {pages} {241102}
  (\bibinfo {year} {2016})}\BibitemShut {NoStop}%
\bibitem [{\citenamefont {Nayak}\ \emph {et~al.}(2008)\citenamefont {Nayak},
  \citenamefont {Simon}, \citenamefont {Stern}, \citenamefont {Freedman},\ and\
  \citenamefont {Das~Sarma}}]{Majorana}%
  \BibitemOpen
  \bibfield  {author} {\bibinfo {author} {\bibfnamefont {C.}~\bibnamefont
  {Nayak}}, \bibinfo {author} {\bibfnamefont {S.~H.}\ \bibnamefont {Simon}},
  \bibinfo {author} {\bibfnamefont {A.}~\bibnamefont {Stern}}, \bibinfo
  {author} {\bibfnamefont {M.}~\bibnamefont {Freedman}}, \ and\ \bibinfo
  {author} {\bibfnamefont {S.}~\bibnamefont {Das~Sarma}},\ }\bibfield  {title}
  {\enquote {\bibinfo {title} {Non-abelian anyons and topological quantum
  computation},}\ }\href {\doibase 10.1103/RevModPhys.80.1083} {\bibfield
  {journal} {\bibinfo  {journal} {Rev. Mod. Phys.}\ }\textbf {\bibinfo {volume}
  {80}},\ \bibinfo {pages} {1083--1159} (\bibinfo {year} {2008})}\BibitemShut
  {NoStop}%
\bibitem [{\citenamefont {Shapourian}, \citenamefont {Wang},\ and\
  \citenamefont {Ryu}(2018)}]{Shapourian2018}%
  \BibitemOpen
  \bibfield  {author} {\bibinfo {author} {\bibfnamefont {H.}~\bibnamefont
  {Shapourian}}, \bibinfo {author} {\bibfnamefont {Y.}~\bibnamefont {Wang}}, \
  and\ \bibinfo {author} {\bibfnamefont {S.}~\bibnamefont {Ryu}},\ }\bibfield
  {title} {\enquote {\bibinfo {title} {Topological crystalline
  superconductivity and second-order topological superconductivity in
  nodal-loop materials},}\ }\href {\doibase 10.1103/PhysRevB.97.094508}
  {\bibfield  {journal} {\bibinfo  {journal} {Phys. Rev. B}\ }\textbf {\bibinfo
  {volume} {97}},\ \bibinfo {pages} {094508} (\bibinfo {year}
  {2018})}\BibitemShut {NoStop}%
\bibitem [{\citenamefont {Qi}\ and\ \citenamefont {Zhang}(2011)}]{Maj_2}%
  \BibitemOpen
  \bibfield  {author} {\bibinfo {author} {\bibfnamefont {X.-L.}\ \bibnamefont
  {Qi}}\ and\ \bibinfo {author} {\bibfnamefont {S.-C.}\ \bibnamefont {Zhang}},\
  }\bibfield  {title} {\enquote {\bibinfo {title} {Topological insulators and
  superconductors},}\ }\href@noop {} {\bibfield  {journal} {\bibinfo  {journal}
  {Rev. Mod. Phys.}\ }\textbf {\bibinfo {volume} {83}},\ \bibinfo {pages}
  {1057--1110} (\bibinfo {year} {2011})}\BibitemShut {NoStop}%
\bibitem [{\citenamefont {Hosur}\ \emph {et~al.}(2011)\citenamefont {Hosur},
  \citenamefont {Ghaemi}, \citenamefont {Mong},\ and\ \citenamefont
  {Vishwanath}}]{Maj_3}%
  \BibitemOpen
  \bibfield  {author} {\bibinfo {author} {\bibfnamefont {P.}~\bibnamefont
  {Hosur}}, \bibinfo {author} {\bibfnamefont {P.}~\bibnamefont {Ghaemi}},
  \bibinfo {author} {\bibfnamefont {R.~S.~K.}\ \bibnamefont {Mong}}, \ and\
  \bibinfo {author} {\bibfnamefont {A.}~\bibnamefont {Vishwanath}},\ }\bibfield
   {title} {\enquote {\bibinfo {title} {Majorana modes at the ends of
  superconductor vortices in doped topological insulators},}\ }\href {\doibase
  10.1103/PhysRevLett.107.097001} {\bibfield  {journal} {\bibinfo  {journal}
  {Phys. Rev. Lett.}\ }\textbf {\bibinfo {volume} {107}},\ \bibinfo {pages}
  {097001} (\bibinfo {year} {2011})}\BibitemShut {NoStop}%
\bibitem [{\citenamefont {Hor}\ \emph {et~al.}(2010)\citenamefont {Hor},
  \citenamefont {Williams}, \citenamefont {Checkelsky}, \citenamefont
  {Roushan}, \citenamefont {Seo}, \citenamefont {Xu}, \citenamefont
  {Zandbergen}, \citenamefont {Yazdani}, \citenamefont {Ong},\ and\
  \citenamefont {Cava}}]{cuxBi2Se3}%
  \BibitemOpen
  \bibfield  {author} {\bibinfo {author} {\bibfnamefont {Y.~S.}\ \bibnamefont
  {Hor}}, \bibinfo {author} {\bibfnamefont {A.~J.}\ \bibnamefont {Williams}},
  \bibinfo {author} {\bibfnamefont {J.~G.}\ \bibnamefont {Checkelsky}},
  \bibinfo {author} {\bibfnamefont {P.}~\bibnamefont {Roushan}}, \bibinfo
  {author} {\bibfnamefont {J.}~\bibnamefont {Seo}}, \bibinfo {author}
  {\bibfnamefont {Q.}~\bibnamefont {Xu}}, \bibinfo {author} {\bibfnamefont
  {H.~W.}\ \bibnamefont {Zandbergen}}, \bibinfo {author} {\bibfnamefont
  {A.}~\bibnamefont {Yazdani}}, \bibinfo {author} {\bibfnamefont {N.~P.}\
  \bibnamefont {Ong}}, \ and\ \bibinfo {author} {\bibfnamefont {R.~J.}\
  \bibnamefont {Cava}},\ }\bibfield  {title} {\enquote {\bibinfo {title}
  {{Superconductivity in ${\mathrm{Cu}}_{x}{\mathrm{Bi}}_{2}{\mathrm{Se}}_{3}$
  and its Implications for Pairing in the Undoped Topological Insulator}},}\
  }\href {\doibase 10.1103/PhysRevLett.104.057001} {\bibfield  {journal}
  {\bibinfo  {journal} {Phys. Rev. Lett.}\ }\textbf {\bibinfo {volume} {104}},\
  \bibinfo {pages} {057001} (\bibinfo {year} {2010})}\BibitemShut {NoStop}%
\bibitem [{\citenamefont {Arpino}\ \emph {et~al.}(2014)\citenamefont {Arpino},
  \citenamefont {Wallace}, \citenamefont {Nie}, \citenamefont {Birol},
  \citenamefont {King}, \citenamefont {Chatterjee}, \citenamefont {Uchida},
  \citenamefont {Koohpayeh}, \citenamefont {Wen}, \citenamefont {Page},
  \citenamefont {Fennie}, \citenamefont {Shen},\ and\ \citenamefont
  {McQueen}}]{Tl5Te3}%
  \BibitemOpen
  \bibfield  {author} {\bibinfo {author} {\bibfnamefont {K.~E.}\ \bibnamefont
  {Arpino}}, \bibinfo {author} {\bibfnamefont {D.~C.}\ \bibnamefont {Wallace}},
  \bibinfo {author} {\bibfnamefont {Y.~F.}\ \bibnamefont {Nie}}, \bibinfo
  {author} {\bibfnamefont {T.}~\bibnamefont {Birol}}, \bibinfo {author}
  {\bibfnamefont {P.~D.~C.}\ \bibnamefont {King}}, \bibinfo {author}
  {\bibfnamefont {S.}~\bibnamefont {Chatterjee}}, \bibinfo {author}
  {\bibfnamefont {M.}~\bibnamefont {Uchida}}, \bibinfo {author} {\bibfnamefont
  {S.~M.}\ \bibnamefont {Koohpayeh}}, \bibinfo {author} {\bibfnamefont {J.-J.}\
  \bibnamefont {Wen}}, \bibinfo {author} {\bibfnamefont {K.}~\bibnamefont
  {Page}}, \bibinfo {author} {\bibfnamefont {C.~J.}\ \bibnamefont {Fennie}},
  \bibinfo {author} {\bibfnamefont {K.~M.}\ \bibnamefont {Shen}}, \ and\
  \bibinfo {author} {\bibfnamefont {T.~M.}\ \bibnamefont {McQueen}},\
  }\bibfield  {title} {\enquote {\bibinfo {title} {{Evidence for Topologically
  Protected Surface States and a Superconducting Phase in
  $[{\mathrm{Tl}}_{4}]({\mathrm{Tl}}_{1\ensuremath{-}x}{\mathrm{Sn}}_{x}){\mathrm{Te}}_{3}$Using
  Photoemission, Specific Heat, and Magnetization Measurements, and Density
  Functional Theory}},}\ }\href {\doibase 10.1103/PhysRevLett.112.017002}
  {\bibfield  {journal} {\bibinfo  {journal} {Phys. Rev. Lett.}\ }\textbf
  {\bibinfo {volume} {112}},\ \bibinfo {pages} {017002} (\bibinfo {year}
  {2014})}\BibitemShut {NoStop}%
\bibitem [{\citenamefont {Aggarwal}\ \emph {et~al.}(2016)\citenamefont
  {Aggarwal}, \citenamefont {Thakur}, \citenamefont {Haque}, \citenamefont
  {Ganguli},\ and\ \citenamefont {Sheet}}]{Cd3As2}%
  \BibitemOpen
  \bibfield  {author} {\bibinfo {author} {\bibfnamefont {A.}~\bibnamefont
  {Aggarwal}, \bibfnamefont {L.and~Gaurav}}, \bibinfo {author} {\bibfnamefont
  {G.~S.}\ \bibnamefont {Thakur}}, \bibinfo {author} {\bibfnamefont
  {Z.}~\bibnamefont {Haque}}, \bibinfo {author} {\bibfnamefont {A.~K.}\
  \bibnamefont {Ganguli}}, \ and\ \bibinfo {author} {\bibfnamefont
  {G.}~\bibnamefont {Sheet}},\ }\bibfield  {title} {\enquote {\bibinfo {title}
  {{Unconventional superconductivity at mesoscopic point contacts on the 3D
  Dirac semimetal \ce{Cd3As2}}},}\ }\href@noop {} {\bibfield  {journal}
  {\bibinfo  {journal} {Nature materials}\ }\textbf {\bibinfo {volume} {15}},\
  \bibinfo {pages} {32} (\bibinfo {year} {2016})}\BibitemShut {NoStop}%
\bibitem [{\citenamefont {Schoop}\ \emph {et~al.}(2015)\citenamefont {Schoop},
  \citenamefont {Xie}, \citenamefont {Chen}, \citenamefont {Gibson},
  \citenamefont {Lapidus}, \citenamefont {Kimchi}, \citenamefont
  {Hirschberger}, \citenamefont {Haldolaarachchige}, \citenamefont {Ali},
  \citenamefont {Belvin}, \citenamefont {Liang}, \citenamefont {Neaton},
  \citenamefont {Ong}, \citenamefont {Vishwanath},\ and\ \citenamefont
  {Cava}}]{Schoop2015Dirac}%
  \BibitemOpen
  \bibfield  {author} {\bibinfo {author} {\bibfnamefont {L.~M.}\ \bibnamefont
  {Schoop}}, \bibinfo {author} {\bibfnamefont {L.~S.}\ \bibnamefont {Xie}},
  \bibinfo {author} {\bibfnamefont {R.}~\bibnamefont {Chen}}, \bibinfo {author}
  {\bibfnamefont {Q.~D.}\ \bibnamefont {Gibson}}, \bibinfo {author}
  {\bibfnamefont {S.~H.}\ \bibnamefont {Lapidus}}, \bibinfo {author}
  {\bibfnamefont {I.}~\bibnamefont {Kimchi}}, \bibinfo {author} {\bibfnamefont
  {M.}~\bibnamefont {Hirschberger}}, \bibinfo {author} {\bibfnamefont
  {N.}~\bibnamefont {Haldolaarachchige}}, \bibinfo {author} {\bibfnamefont
  {M.~N.}\ \bibnamefont {Ali}}, \bibinfo {author} {\bibfnamefont {C.~A.}\
  \bibnamefont {Belvin}}, \bibinfo {author} {\bibfnamefont {T.}~\bibnamefont
  {Liang}}, \bibinfo {author} {\bibfnamefont {J.~B.}\ \bibnamefont {Neaton}},
  \bibinfo {author} {\bibfnamefont {N.~P.}\ \bibnamefont {Ong}}, \bibinfo
  {author} {\bibfnamefont {A.}~\bibnamefont {Vishwanath}}, \ and\ \bibinfo
  {author} {\bibfnamefont {R.~J.}\ \bibnamefont {Cava}},\ }\bibfield  {title}
  {\enquote {\bibinfo {title} {Dirac metal to topological metal transition at a
  structural phase change in ${\mathrm{au}}_{2}\mathrm{Pb}$ and prediction of
  ${\mathbb{z}}_{2}$ topology for the superconductor},}\ }\href {\doibase
  10.1103/PhysRevB.91.214517} {\bibfield  {journal} {\bibinfo  {journal} {Phys.
  Rev. B}\ }\textbf {\bibinfo {volume} {91}},\ \bibinfo {pages} {214517}
  (\bibinfo {year} {2015})}\BibitemShut {NoStop}%
\bibitem [{\citenamefont {Yuan}\ \emph {et~al.}(2019)\citenamefont {Yuan},
  \citenamefont {Pan}, \citenamefont {Wang}, \citenamefont {Fang},
  \citenamefont {Song}, \citenamefont {Wang}, \citenamefont {He}, \citenamefont
  {Ma}, \citenamefont {Zhang}, \citenamefont {Huang}, \citenamefont {Li},\ and\
  \citenamefont {Xue}}]{WS2_1}%
  \BibitemOpen
  \bibfield  {author} {\bibinfo {author} {\bibfnamefont {Y.}~\bibnamefont
  {Yuan}}, \bibinfo {author} {\bibfnamefont {J.}~\bibnamefont {Pan}}, \bibinfo
  {author} {\bibfnamefont {X.}~\bibnamefont {Wang}}, \bibinfo {author}
  {\bibfnamefont {Y.}~\bibnamefont {Fang}}, \bibinfo {author} {\bibfnamefont
  {C.}~\bibnamefont {Song}}, \bibinfo {author} {\bibfnamefont {L.}~\bibnamefont
  {Wang}}, \bibinfo {author} {\bibfnamefont {K.}~\bibnamefont {He}}, \bibinfo
  {author} {\bibfnamefont {X.}~\bibnamefont {Ma}}, \bibinfo {author}
  {\bibfnamefont {H.}~\bibnamefont {Zhang}}, \bibinfo {author} {\bibfnamefont
  {F.}~\bibnamefont {Huang}}, \bibinfo {author} {\bibfnamefont
  {W.}~\bibnamefont {Li}}, \ and\ \bibinfo {author} {\bibfnamefont {Q.-K.}\
  \bibnamefont {Xue}},\ }\bibfield  {title} {\enquote {\bibinfo {title}
  {{Evidence of anisotropic Majorana bound states in 2M-\ce{WS2}}},}\ }\href
  {https://doi.org/10.1038/s41567-019-0576-7} {\bibfield  {journal} {\bibinfo
  {journal} {Nature Physics}\ } (\bibinfo {year} {2019})}\BibitemShut {NoStop}%
\bibitem [{\citenamefont {{Guguchia}}\ \emph {et~al.}(2019)\citenamefont
  {{Guguchia}}, \citenamefont {{Gawryluk}}, \citenamefont {{Brzezinska}},
  \citenamefont {{Tsirkin}}, \citenamefont {{Khasanov}}, \citenamefont
  {{Pomjakushina}}, \citenamefont {{von Rohr}}, \citenamefont {{Verezhak}},
  \citenamefont {{Hasan}}, \citenamefont {{Neupert}}, \citenamefont
  {{Luetkens}},\ and\ \citenamefont {{Amato}}}]{WS2_2}%
  \BibitemOpen
  \bibfield  {author} {\bibinfo {author} {\bibfnamefont {Z.}~\bibnamefont
  {{Guguchia}}}, \bibinfo {author} {\bibfnamefont {D.}~\bibnamefont
  {{Gawryluk}}}, \bibinfo {author} {\bibfnamefont {M.}~\bibnamefont
  {{Brzezinska}}}, \bibinfo {author} {\bibfnamefont {S.~S.}\ \bibnamefont
  {{Tsirkin}}}, \bibinfo {author} {\bibfnamefont {R.}~\bibnamefont
  {{Khasanov}}}, \bibinfo {author} {\bibfnamefont {E.}~\bibnamefont
  {{Pomjakushina}}}, \bibinfo {author} {\bibfnamefont {F.~O.}\ \bibnamefont
  {{von Rohr}}}, \bibinfo {author} {\bibfnamefont {J.}~\bibnamefont
  {{Verezhak}}}, \bibinfo {author} {\bibfnamefont {M.~Z.}\ \bibnamefont
  {{Hasan}}}, \bibinfo {author} {\bibfnamefont {T.}~\bibnamefont {{Neupert}}},
  \bibinfo {author} {\bibfnamefont {H.}~\bibnamefont {{Luetkens}}}, \ and\
  \bibinfo {author} {\bibfnamefont {A.}~\bibnamefont {{Amato}}},\ }\bibfield
  {title} {\enquote {\bibinfo {title} {{Nodeless Superconductivity and its
  Evolution with Pressure in the Layered Dirac Semimetal 2M-\ce{WS2}}},}\
  }\href@noop {} {\bibfield  {journal} {\bibinfo  {journal} {arXiv preprint
  arXiv:1903.10612}\ } (\bibinfo {year} {2019})}\BibitemShut {NoStop}%
\bibitem [{\citenamefont {Qi}\ \emph {et~al.}(2016)\citenamefont {Qi},
  \citenamefont {Naumov}, \citenamefont {Ali}, \citenamefont {Rajamathi},
  \citenamefont {Schnelle}, \citenamefont {Barkalov}, \citenamefont {Hanfland},
  \citenamefont {Wu}, \citenamefont {Shekhar}, \citenamefont {Sun},
  \citenamefont {Süß}, \citenamefont {Schmidt}, \citenamefont {Schwarz},
  \citenamefont {Pippel}, \citenamefont {Werner}, \citenamefont {Hillebrand},
  \citenamefont {Förster}, \citenamefont {Kampert}, \citenamefont {Parkin},
  \citenamefont {Cava}, \citenamefont {Felser}, \citenamefont {Yan},\ and\
  \citenamefont {Medvedev}}]{MoTe2_1}%
  \BibitemOpen
  \bibfield  {author} {\bibinfo {author} {\bibfnamefont {Y.}~\bibnamefont
  {Qi}}, \bibinfo {author} {\bibfnamefont {P.~G.}\ \bibnamefont {Naumov}},
  \bibinfo {author} {\bibfnamefont {M.~N.}\ \bibnamefont {Ali}}, \bibinfo
  {author} {\bibfnamefont {C.~R.}\ \bibnamefont {Rajamathi}}, \bibinfo {author}
  {\bibfnamefont {W.}~\bibnamefont {Schnelle}}, \bibinfo {author}
  {\bibfnamefont {O.}~\bibnamefont {Barkalov}}, \bibinfo {author}
  {\bibfnamefont {M.}~\bibnamefont {Hanfland}}, \bibinfo {author}
  {\bibfnamefont {S.-C.}\ \bibnamefont {Wu}}, \bibinfo {author} {\bibfnamefont
  {C.}~\bibnamefont {Shekhar}}, \bibinfo {author} {\bibfnamefont
  {Y.}~\bibnamefont {Sun}}, \bibinfo {author} {\bibfnamefont {V.}~\bibnamefont
  {Süß}}, \bibinfo {author} {\bibfnamefont {M.}~\bibnamefont {Schmidt}},
  \bibinfo {author} {\bibfnamefont {U.}~\bibnamefont {Schwarz}}, \bibinfo
  {author} {\bibfnamefont {E.}~\bibnamefont {Pippel}}, \bibinfo {author}
  {\bibfnamefont {P.}~\bibnamefont {Werner}}, \bibinfo {author} {\bibfnamefont
  {R.}~\bibnamefont {Hillebrand}}, \bibinfo {author} {\bibfnamefont
  {T.}~\bibnamefont {Förster}}, \bibinfo {author} {\bibfnamefont
  {E.}~\bibnamefont {Kampert}}, \bibinfo {author} {\bibfnamefont
  {S.}~\bibnamefont {Parkin}}, \bibinfo {author} {\bibfnamefont {R.~J.}\
  \bibnamefont {Cava}}, \bibinfo {author} {\bibfnamefont {C.}~\bibnamefont
  {Felser}}, \bibinfo {author} {\bibfnamefont {B.}~\bibnamefont {Yan}}, \ and\
  \bibinfo {author} {\bibfnamefont {S.~A.}\ \bibnamefont {Medvedev}},\
  }\bibfield  {title} {\enquote {\bibinfo {title} {Superconductivity in weyl
  semimetal candidate \ce{MoTe2}},}\ }\href
  {https://doi.org/10.1038/ncomms11038} {\bibfield  {journal} {\bibinfo
  {journal} {Nature Communications}\ }\textbf {\bibinfo {volume} {7}},\
  \bibinfo {pages} {11038} (\bibinfo {year} {2016})}\BibitemShut {NoStop}%
\bibitem [{\citenamefont {Guguchia}\ \emph {et~al.}(2017)\citenamefont
  {Guguchia}, \citenamefont {von Rohr}, \citenamefont {Shermadini},
  \citenamefont {Lee}, \citenamefont {Banerjee}, \citenamefont {Wieteska},
  \citenamefont {Marianetti}, \citenamefont {Frandsen}, \citenamefont
  {Luetkens}, \citenamefont {Gong}, \citenamefont {Cheung}, \citenamefont
  {Baines}, \citenamefont {Shengelaya}, \citenamefont {Taniashvili},
  \citenamefont {Pasupathy}, \citenamefont {Morenzoni}, \citenamefont
  {Billinge}, \citenamefont {Amato}, \citenamefont {Cava}, \citenamefont
  {Khasanov},\ and\ \citenamefont {Uemura}}]{MoTe2_2}%
  \BibitemOpen
  \bibfield  {author} {\bibinfo {author} {\bibfnamefont {Z.}~\bibnamefont
  {Guguchia}}, \bibinfo {author} {\bibfnamefont {F.}~\bibnamefont {von Rohr}},
  \bibinfo {author} {\bibfnamefont {Z.}~\bibnamefont {Shermadini}}, \bibinfo
  {author} {\bibfnamefont {A.~T.}\ \bibnamefont {Lee}}, \bibinfo {author}
  {\bibfnamefont {S.}~\bibnamefont {Banerjee}}, \bibinfo {author}
  {\bibfnamefont {A.~R.}\ \bibnamefont {Wieteska}}, \bibinfo {author}
  {\bibfnamefont {C.~A.}\ \bibnamefont {Marianetti}}, \bibinfo {author}
  {\bibfnamefont {B.~A.}\ \bibnamefont {Frandsen}}, \bibinfo {author}
  {\bibfnamefont {H.}~\bibnamefont {Luetkens}}, \bibinfo {author}
  {\bibfnamefont {Z.}~\bibnamefont {Gong}}, \bibinfo {author} {\bibfnamefont
  {S.~C.}\ \bibnamefont {Cheung}}, \bibinfo {author} {\bibfnamefont
  {C.}~\bibnamefont {Baines}}, \bibinfo {author} {\bibfnamefont
  {A.}~\bibnamefont {Shengelaya}}, \bibinfo {author} {\bibfnamefont
  {G.}~\bibnamefont {Taniashvili}}, \bibinfo {author} {\bibfnamefont {A.~N.}\
  \bibnamefont {Pasupathy}}, \bibinfo {author} {\bibfnamefont {E.}~\bibnamefont
  {Morenzoni}}, \bibinfo {author} {\bibfnamefont {S.~J.~L.}\ \bibnamefont
  {Billinge}}, \bibinfo {author} {\bibfnamefont {A.}~\bibnamefont {Amato}},
  \bibinfo {author} {\bibfnamefont {R.~J.}\ \bibnamefont {Cava}}, \bibinfo
  {author} {\bibfnamefont {R.}~\bibnamefont {Khasanov}}, \ and\ \bibinfo
  {author} {\bibfnamefont {Y.~J.}\ \bibnamefont {Uemura}},\ }\bibfield  {title}
  {\enquote {\bibinfo {title} {{Signatures of the topological s$^{+-}$
  superconducting order parameter in the type-II Weyl semimetal T$_{\rm
  d}$-\ce{MoTe2}}},}\ }\href {https://doi.org/10.1038/s41467-017-01066-6}
  {\bibfield  {journal} {\bibinfo  {journal} {Nature Communications}\ }\textbf
  {\bibinfo {volume} {8}},\ \bibinfo {pages} {1082} (\bibinfo {year}
  {2017})}\BibitemShut {NoStop}%
\bibitem [{\citenamefont {Bian}\ \emph {et~al.}(2016)\citenamefont {Bian},
  \citenamefont {Chang}, \citenamefont {Sankar}, \citenamefont {Xu},
  \citenamefont {Zheng}, \citenamefont {Neupert}, \citenamefont {Chiu},
  \citenamefont {Huang}, \citenamefont {Chang}, \citenamefont {Belopolski},
  \citenamefont {Sanchez}, \citenamefont {Neupane}, \citenamefont {Alidoust},
  \citenamefont {Liu}, \citenamefont {Wang}, \citenamefont {Lee}, \citenamefont
  {Jeng}, \citenamefont {Zhang}, \citenamefont {Yuan}, \citenamefont {Jia},
  \citenamefont {Bansil}, \citenamefont {Chou}, \citenamefont {Lin},\ and\
  \citenamefont {Hasan}}]{PbTaSe2}%
  \BibitemOpen
  \bibfield  {author} {\bibinfo {author} {\bibfnamefont {G.}~\bibnamefont
  {Bian}}, \bibinfo {author} {\bibfnamefont {T.-R.}\ \bibnamefont {Chang}},
  \bibinfo {author} {\bibfnamefont {R.}~\bibnamefont {Sankar}}, \bibinfo
  {author} {\bibfnamefont {S.-Y.}\ \bibnamefont {Xu}}, \bibinfo {author}
  {\bibfnamefont {H.}~\bibnamefont {Zheng}}, \bibinfo {author} {\bibfnamefont
  {T.}~\bibnamefont {Neupert}}, \bibinfo {author} {\bibfnamefont {C.-K.}\
  \bibnamefont {Chiu}}, \bibinfo {author} {\bibfnamefont {S.-M.}\ \bibnamefont
  {Huang}}, \bibinfo {author} {\bibfnamefont {G.}~\bibnamefont {Chang}},
  \bibinfo {author} {\bibfnamefont {I.}~\bibnamefont {Belopolski}}, \bibinfo
  {author} {\bibfnamefont {D.~S.}\ \bibnamefont {Sanchez}}, \bibinfo {author}
  {\bibfnamefont {M.}~\bibnamefont {Neupane}}, \bibinfo {author} {\bibfnamefont
  {N.}~\bibnamefont {Alidoust}}, \bibinfo {author} {\bibfnamefont
  {C.}~\bibnamefont {Liu}}, \bibinfo {author} {\bibfnamefont {B.}~\bibnamefont
  {Wang}}, \bibinfo {author} {\bibfnamefont {C.-C.}\ \bibnamefont {Lee}},
  \bibinfo {author} {\bibfnamefont {H.-T.}\ \bibnamefont {Jeng}}, \bibinfo
  {author} {\bibfnamefont {C.}~\bibnamefont {Zhang}}, \bibinfo {author}
  {\bibfnamefont {Z.}~\bibnamefont {Yuan}}, \bibinfo {author} {\bibfnamefont
  {S.}~\bibnamefont {Jia}}, \bibinfo {author} {\bibfnamefont {A.}~\bibnamefont
  {Bansil}}, \bibinfo {author} {\bibfnamefont {F.}~\bibnamefont {Chou}},
  \bibinfo {author} {\bibfnamefont {H.}~\bibnamefont {Lin}}, \ and\ \bibinfo
  {author} {\bibfnamefont {M.~Z.}\ \bibnamefont {Hasan}},\ }\bibfield  {title}
  {\enquote {\bibinfo {title} {Topological nodal-line fermions in spin-orbit
  metal \ce{PbTaSe2}},}\ }\href@noop {} {\bibfield  {journal} {\bibinfo
  {journal} {Nature Communications}\ }\textbf {\bibinfo {volume} {7}},\
  \bibinfo {pages} {10556} (\bibinfo {year} {2016})}\BibitemShut {NoStop}%
\bibitem [{\citenamefont {Fang}\ \emph {et~al.}(2016)\citenamefont {Fang},
  \citenamefont {Weng}, \citenamefont {Dai},\ and\ \citenamefont
  {Fang}}]{nodal_line_SC}%
  \BibitemOpen
  \bibfield  {author} {\bibinfo {author} {\bibfnamefont {C.}~\bibnamefont
  {Fang}}, \bibinfo {author} {\bibfnamefont {H.}~\bibnamefont {Weng}}, \bibinfo
  {author} {\bibfnamefont {X.}~\bibnamefont {Dai}}, \ and\ \bibinfo {author}
  {\bibfnamefont {Z.}~\bibnamefont {Fang}},\ }\bibfield  {title} {\enquote
  {\bibinfo {title} {Topological nodal line semimetals},}\ }\href@noop {}
  {\bibfield  {journal} {\bibinfo  {journal} {Chin. Phys. B}\ }\textbf
  {\bibinfo {volume} {25}},\ \bibinfo {pages} {117106} (\bibinfo {year}
  {2016})}\BibitemShut {NoStop}%
\bibitem [{\citenamefont {Guan}\ \emph {et~al.}(2016)\citenamefont {Guan},
  \citenamefont {Chen}, \citenamefont {Chu}, \citenamefont {Sankar},
  \citenamefont {C.}, \citenamefont {Jeng}, \citenamefont {Chang},\ and\
  \citenamefont {Chuang}}]{PbTaSe2_2}%
  \BibitemOpen
  \bibfield  {author} {\bibinfo {author} {\bibfnamefont {S.-Y.}\ \bibnamefont
  {Guan}}, \bibinfo {author} {\bibfnamefont {P.-J.}\ \bibnamefont {Chen}},
  \bibinfo {author} {\bibfnamefont {M.-W.}\ \bibnamefont {Chu}}, \bibinfo
  {author} {\bibfnamefont {R.}~\bibnamefont {Sankar}}, \bibinfo {author}
  {\bibfnamefont {F.}~\bibnamefont {C.}}, \bibinfo {author} {\bibfnamefont
  {H.-T.}\ \bibnamefont {Jeng}}, \bibinfo {author} {\bibfnamefont {C.-S.}\
  \bibnamefont {Chang}}, \ and\ \bibinfo {author} {\bibfnamefont {T.-M.}\
  \bibnamefont {Chuang}},\ }\bibfield  {title} {\enquote {\bibinfo {title}
  {{Superconducting topological surface states in the noncentrosymmetric bulk
  superconductor \ce{PbTaSe2}}},}\ }\href {\doibase 10.1126/sciadv.1600894}
  {\bibfield  {journal} {\bibinfo  {journal} {Science Advances}\ }\textbf
  {\bibinfo {volume} {2}} (\bibinfo {year} {2016}),\
  10.1126/sciadv.1600894}\BibitemShut {NoStop}%
\bibitem [{\citenamefont {Ali}\ \emph {et~al.}(2014)\citenamefont {Ali},
  \citenamefont {Gibson}, \citenamefont {Klimczuk},\ and\ \citenamefont
  {Cava}}]{PbTaSe2_3}%
  \BibitemOpen
  \bibfield  {author} {\bibinfo {author} {\bibfnamefont {M.~N.}\ \bibnamefont
  {Ali}}, \bibinfo {author} {\bibfnamefont {Q.~D.}\ \bibnamefont {Gibson}},
  \bibinfo {author} {\bibfnamefont {T.}~\bibnamefont {Klimczuk}}, \ and\
  \bibinfo {author} {\bibfnamefont {R.~J.}\ \bibnamefont {Cava}},\ }\bibfield
  {title} {\enquote {\bibinfo {title} {{Noncentrosymmetric superconductor with
  a bulk three-dimensional Dirac cone gapped by strong spin-orbit coupling}},}\
  }\href@noop {} {\bibfield  {journal} {\bibinfo  {journal} {Phys. Rev. B}\
  }\textbf {\bibinfo {volume} {89}},\ \bibinfo {pages} {020505} (\bibinfo
  {year} {2014})}\BibitemShut {NoStop}%
\bibitem [{\citenamefont {Westerhaus}\ and\ \citenamefont
  {Schuster}(1979)}]{NaAlSi_structure}%
  \BibitemOpen
  \bibfield  {author} {\bibinfo {author} {\bibfnamefont {W.}~\bibnamefont
  {Westerhaus}}\ and\ \bibinfo {author} {\bibfnamefont {H.-U.}\ \bibnamefont
  {Schuster}},\ }\bibfield  {title} {\enquote {\bibinfo {title} {{Darstellung
  und Struktur von NaAlSi und NaAlGe}},}\ }\href@noop {} {\bibfield  {journal}
  {\bibinfo  {journal} {Zeitschrift f\"ur Naturforschung B}\ }\textbf {\bibinfo
  {volume} {34}},\ \bibinfo {pages} {352} (\bibinfo {year} {1979})}\BibitemShut
  {NoStop}%
\bibitem [{\citenamefont {Kuroiwa}\ \emph {et~al.}(2007)\citenamefont
  {Kuroiwa}, \citenamefont {Kawashima}, \citenamefont {Kinoshita},
  \citenamefont {Okabe},\ and\ \citenamefont {Akimitsu}}]{NaAlSi_2}%
  \BibitemOpen
  \bibfield  {author} {\bibinfo {author} {\bibfnamefont {S.}~\bibnamefont
  {Kuroiwa}}, \bibinfo {author} {\bibfnamefont {H.}~\bibnamefont {Kawashima}},
  \bibinfo {author} {\bibfnamefont {H.}~\bibnamefont {Kinoshita}}, \bibinfo
  {author} {\bibfnamefont {H.}~\bibnamefont {Okabe}}, \ and\ \bibinfo {author}
  {\bibfnamefont {J.}~\bibnamefont {Akimitsu}},\ }\bibfield  {title} {\enquote
  {\bibinfo {title} {{Superconductivity in ternary silicide NaAlSi with layered
  diamond-like structure}},}\ }\href {\doibase
  https://doi.org/10.1016/j.physc.2007.04.232} {\bibfield  {journal} {\bibinfo
  {journal} {Physica C: Superconductivity}\ }\textbf {\bibinfo {volume}
  {466}},\ \bibinfo {pages} {11 -- 15} (\bibinfo {year} {2007})}\BibitemShut
  {NoStop}%
\bibitem [{\citenamefont {Rhee}\ \emph {et~al.}(2010)\citenamefont {Rhee},
  \citenamefont {Banerjee}, \citenamefont {Ylvisaker},\ and\ \citenamefont
  {Pickett}}]{NaAlSi_Pickett}%
  \BibitemOpen
  \bibfield  {author} {\bibinfo {author} {\bibfnamefont {H.~B.}\ \bibnamefont
  {Rhee}}, \bibinfo {author} {\bibfnamefont {S.}~\bibnamefont {Banerjee}},
  \bibinfo {author} {\bibfnamefont {E.~R.}\ \bibnamefont {Ylvisaker}}, \ and\
  \bibinfo {author} {\bibfnamefont {W.~E.}\ \bibnamefont {Pickett}},\
  }\bibfield  {title} {\enquote {\bibinfo {title} {{NaAlSi: Self-doped
  semimetallic superconductor with free electrons and covalent holes}},}\
  }\href {\doibase 10.1103/PhysRevB.81.245114} {\bibfield  {journal} {\bibinfo
  {journal} {Phys. Rev. B}\ }\textbf {\bibinfo {volume} {81}},\ \bibinfo
  {pages} {245114} (\bibinfo {year} {2010})}\BibitemShut {NoStop}%
\bibitem [{\citenamefont {Schoop}\ \emph {et~al.}(2012)\citenamefont {Schoop},
  \citenamefont {M\"uchler}, \citenamefont {Schmitt}, \citenamefont
  {Ksenofontov}, \citenamefont {Medvedev}, \citenamefont {Nuss}, \citenamefont
  {Casper}, \citenamefont {Jansen}, \citenamefont {Cava},\ and\ \citenamefont
  {Felser}}]{Schoop2012Effect}%
  \BibitemOpen
  \bibfield  {author} {\bibinfo {author} {\bibfnamefont {L.}~\bibnamefont
  {Schoop}}, \bibinfo {author} {\bibfnamefont {L.}~\bibnamefont {M\"uchler}},
  \bibinfo {author} {\bibfnamefont {J.}~\bibnamefont {Schmitt}}, \bibinfo
  {author} {\bibfnamefont {V.}~\bibnamefont {Ksenofontov}}, \bibinfo {author}
  {\bibfnamefont {S.}~\bibnamefont {Medvedev}}, \bibinfo {author}
  {\bibfnamefont {J.}~\bibnamefont {Nuss}}, \bibinfo {author} {\bibfnamefont
  {F.}~\bibnamefont {Casper}}, \bibinfo {author} {\bibfnamefont
  {M.}~\bibnamefont {Jansen}}, \bibinfo {author} {\bibfnamefont {R.~J.}\
  \bibnamefont {Cava}}, \ and\ \bibinfo {author} {\bibfnamefont
  {C.}~\bibnamefont {Felser}},\ }\bibfield  {title} {\enquote {\bibinfo {title}
  {Effect of pressure on superconductivity in \ce{NaAlSi}},}\ }\href {\doibase
  10.1103/PhysRevB.86.174522} {\bibfield  {journal} {\bibinfo  {journal} {Phys.
  Rev. B}\ }\textbf {\bibinfo {volume} {86}},\ \bibinfo {pages} {174522}
  (\bibinfo {year} {2012})}\BibitemShut {NoStop}%
\bibitem [{\citenamefont {Kresse}\ and\ \citenamefont
  {Furthm{\"u}ller}(1996)}]{VASP}%
  \BibitemOpen
  \bibfield  {author} {\bibinfo {author} {\bibfnamefont {G.}~\bibnamefont
  {Kresse}}\ and\ \bibinfo {author} {\bibfnamefont {J.}~\bibnamefont
  {Furthm{\"u}ller}},\ }\bibfield  {title} {\enquote {\bibinfo {title}
  {Efficiency of ab-initio total energy calculations for metals and
  semiconductors using a plane-wave basis set},}\ }\href@noop {} {\bibfield
  {journal} {\bibinfo  {journal} {Comput. Mat. Sci.}\ }\textbf {\bibinfo
  {volume} {6}},\ \bibinfo {pages} {15--50} (\bibinfo {year}
  {1996})}\BibitemShut {NoStop}%
\bibitem [{\citenamefont {Wu}\ \emph {et~al.}(2018)\citenamefont {Wu},
  \citenamefont {Zhang}, \citenamefont {Song}, \citenamefont {Troyer},\ and\
  \citenamefont {Soluyanov}}]{wu2018wanniertools}%
  \BibitemOpen
  \bibfield  {author} {\bibinfo {author} {\bibfnamefont {Q.}~\bibnamefont
  {Wu}}, \bibinfo {author} {\bibfnamefont {S.}~\bibnamefont {Zhang}}, \bibinfo
  {author} {\bibfnamefont {H.-F.}\ \bibnamefont {Song}}, \bibinfo {author}
  {\bibfnamefont {M.}~\bibnamefont {Troyer}}, \ and\ \bibinfo {author}
  {\bibfnamefont {A.~A.}\ \bibnamefont {Soluyanov}},\ }\bibfield  {title}
  {\enquote {\bibinfo {title} {Wanniertools: An open-source software package
  for novel topological materials},}\ }\href@noop {} {\bibfield  {journal}
  {\bibinfo  {journal} {Comput. Phys. Commun.}\ }\textbf {\bibinfo {volume}
  {224}},\ \bibinfo {pages} {405--416} (\bibinfo {year} {2018})}\BibitemShut
  {NoStop}%
\bibitem [{\citenamefont {Suter}\ and\ \citenamefont {Wojek}(2012)}]{MUSR}%
  \BibitemOpen
  \bibfield  {author} {\bibinfo {author} {\bibfnamefont {A.}~\bibnamefont
  {Suter}}\ and\ \bibinfo {author} {\bibfnamefont {B.}~\bibnamefont {Wojek}},\
  }\bibfield  {title} {\enquote {\bibinfo {title} {{MUSRFIT: A Free
  Platform-Independent Framework for $\mu$SR Data Analysis}},}\ }\href@noop {}
  {\bibfield  {journal} {\bibinfo  {journal} {Physics Procedia}\ }\textbf
  {\bibinfo {volume} {30}},\ \bibinfo {pages} {69 -- 73} (\bibinfo {year}
  {2012})}\BibitemShut {NoStop}%
\bibitem [{\citenamefont {Schoop}\ \emph {et~al.}(2016)\citenamefont {Schoop},
  \citenamefont {Ali}, \citenamefont {Stra{\ss}er}, \citenamefont {Topp},
  \citenamefont {Varykhalov}, \citenamefont {Marchenko}, \citenamefont
  {Duppel}, \citenamefont {Parkin}, \citenamefont {Lotsch},\ and\ \citenamefont
  {Ast}}]{schoop2016dirac}%
  \BibitemOpen
  \bibfield  {author} {\bibinfo {author} {\bibfnamefont {L.~M.}\ \bibnamefont
  {Schoop}}, \bibinfo {author} {\bibfnamefont {M.~N.}\ \bibnamefont {Ali}},
  \bibinfo {author} {\bibfnamefont {C.}~\bibnamefont {Stra{\ss}er}}, \bibinfo
  {author} {\bibfnamefont {A.}~\bibnamefont {Topp}}, \bibinfo {author}
  {\bibfnamefont {A.}~\bibnamefont {Varykhalov}}, \bibinfo {author}
  {\bibfnamefont {D.}~\bibnamefont {Marchenko}}, \bibinfo {author}
  {\bibfnamefont {V.}~\bibnamefont {Duppel}}, \bibinfo {author} {\bibfnamefont
  {S.~S.}\ \bibnamefont {Parkin}}, \bibinfo {author} {\bibfnamefont {B.~V.}\
  \bibnamefont {Lotsch}}, \ and\ \bibinfo {author} {\bibfnamefont {C.~R.}\
  \bibnamefont {Ast}},\ }\bibfield  {title} {\enquote {\bibinfo {title} {{Dirac
  cone protected by non-symmorphic symmetry and three-dimen\-sional Dirac line
  node in ZrSiS}},}\ }\href@noop {} {\bibfield  {journal} {\bibinfo  {journal}
  {Nat. Commun.}\ }\textbf {\bibinfo {volume} {7}},\ \bibinfo {pages} {11696}
  (\bibinfo {year} {2016})}\BibitemShut {NoStop}%
\bibitem [{\citenamefont {Pezzini}\ \emph {et~al.}(2018)\citenamefont
  {Pezzini}, \citenamefont {Van~Delft}, \citenamefont {Schoop}, \citenamefont
  {Lotsch}, \citenamefont {Carrington}, \citenamefont {Katsnelson},
  \citenamefont {Hussey},\ and\ \citenamefont
  {Wiedmann}}]{pezzini2018unconventional}%
  \BibitemOpen
  \bibfield  {author} {\bibinfo {author} {\bibfnamefont {S.}~\bibnamefont
  {Pezzini}}, \bibinfo {author} {\bibfnamefont {M.~R.}\ \bibnamefont
  {Van~Delft}}, \bibinfo {author} {\bibfnamefont {L.~M.}\ \bibnamefont
  {Schoop}}, \bibinfo {author} {\bibfnamefont {B.~V.}\ \bibnamefont {Lotsch}},
  \bibinfo {author} {\bibfnamefont {A.}~\bibnamefont {Carrington}}, \bibinfo
  {author} {\bibfnamefont {M.~I.}\ \bibnamefont {Katsnelson}}, \bibinfo
  {author} {\bibfnamefont {N.~E.}\ \bibnamefont {Hussey}}, \ and\ \bibinfo
  {author} {\bibfnamefont {S.}~\bibnamefont {Wiedmann}},\ }\bibfield  {title}
  {\enquote {\bibinfo {title} {{Unconventional mass enhancement around the
  Dirac nodal loop in ZrSiS}},}\ }\href@noop {} {\bibfield  {journal} {\bibinfo
   {journal} {Nat. Phys.}\ }\textbf {\bibinfo {volume} {14}},\ \bibinfo {pages}
  {178} (\bibinfo {year} {2018})}\BibitemShut {NoStop}%
\bibitem [{\citenamefont {Zhang}\ \emph {et~al.}(2018)\citenamefont {Zhang},
  \citenamefont {Yaji}, \citenamefont {Hashimoto}, \citenamefont {Ota},
  \citenamefont {Kondo}, \citenamefont {Okazaki}, \citenamefont {Wang},
  \citenamefont {Wen}, \citenamefont {Gu}, \citenamefont {Ding} \emph
  {et~al.}}]{zhang2018observation}%
  \BibitemOpen
  \bibfield  {author} {\bibinfo {author} {\bibfnamefont {P.}~\bibnamefont
  {Zhang}}, \bibinfo {author} {\bibfnamefont {K.}~\bibnamefont {Yaji}},
  \bibinfo {author} {\bibfnamefont {T.}~\bibnamefont {Hashimoto}}, \bibinfo
  {author} {\bibfnamefont {Y.}~\bibnamefont {Ota}}, \bibinfo {author}
  {\bibfnamefont {T.}~\bibnamefont {Kondo}}, \bibinfo {author} {\bibfnamefont
  {K.}~\bibnamefont {Okazaki}}, \bibinfo {author} {\bibfnamefont
  {Z.}~\bibnamefont {Wang}}, \bibinfo {author} {\bibfnamefont {J.}~\bibnamefont
  {Wen}}, \bibinfo {author} {\bibfnamefont {G.}~\bibnamefont {Gu}}, \bibinfo
  {author} {\bibfnamefont {H.}~\bibnamefont {Ding}},  \emph {et~al.},\
  }\bibfield  {title} {\enquote {\bibinfo {title} {Observation of topological
  superconductivity on the surface of an iron-based superconductor},}\
  }\href@noop {} {\bibfield  {journal} {\bibinfo  {journal} {Science}\ }\textbf
  {\bibinfo {volume} {360}},\ \bibinfo {pages} {182--186} (\bibinfo {year}
  {2018})}\BibitemShut {NoStop}%
\bibitem [{\citenamefont {Kubo}\ and\ \citenamefont {Toyabe}(1967)}]{Toyabe}%
  \BibitemOpen
  \bibfield  {author} {\bibinfo {author} {\bibfnamefont {R.}~\bibnamefont
  {Kubo}}\ and\ \bibinfo {author} {\bibfnamefont {T.}~\bibnamefont {Toyabe}},\
  }\bibfield  {title} {\enquote {\bibinfo {title} {Magnetic resonance and
  relaxation},}\ }\href@noop {} {\bibfield  {journal} {\bibinfo  {journal}
  {North-Holland, Amsterdam}\ }\textbf {\bibinfo {volume} {810}} (\bibinfo
  {year} {1967})}\BibitemShut {NoStop}%
\bibitem [{\citenamefont {Luke}\ \emph {et~al.}(1998)\citenamefont {Luke},
  \citenamefont {Fudamoto}, \citenamefont {Kojima}, \citenamefont {Larkin},
  \citenamefont {Merrin}, \citenamefont {Nachumi}, \citenamefont {Uemura},
  \citenamefont {Maeno}, \citenamefont {Mao}, \citenamefont {Mori} \emph
  {et~al.}}]{LukeTRS}%
  \BibitemOpen
  \bibfield  {author} {\bibinfo {author} {\bibfnamefont {G.~M.}\ \bibnamefont
  {Luke}}, \bibinfo {author} {\bibfnamefont {Y.}~\bibnamefont {Fudamoto}},
  \bibinfo {author} {\bibfnamefont {K.}~\bibnamefont {Kojima}}, \bibinfo
  {author} {\bibfnamefont {M.}~\bibnamefont {Larkin}}, \bibinfo {author}
  {\bibfnamefont {J.}~\bibnamefont {Merrin}}, \bibinfo {author} {\bibfnamefont
  {B.}~\bibnamefont {Nachumi}}, \bibinfo {author} {\bibfnamefont
  {Y.}~\bibnamefont {Uemura}}, \bibinfo {author} {\bibfnamefont
  {Y.}~\bibnamefont {Maeno}}, \bibinfo {author} {\bibfnamefont
  {Z.}~\bibnamefont {Mao}}, \bibinfo {author} {\bibfnamefont {Y.}~\bibnamefont
  {Mori}},  \emph {et~al.},\ }\bibfield  {title} {\enquote {\bibinfo {title}
  {Time-reversal symmetry-breaking superconductivity in {Sr$_2$RuO$_4$}},}\
  }\href@noop {} {\bibfield  {journal} {\bibinfo  {journal} {Nature}\ }\textbf
  {\bibinfo {volume} {394}},\ \bibinfo {pages} {558} (\bibinfo {year}
  {1998})}\BibitemShut {NoStop}%
\bibitem [{\citenamefont {Hillier}, \citenamefont {Quintanilla},\ and\
  \citenamefont {Cywinski}(2009)}]{HillierTRS}%
  \BibitemOpen
  \bibfield  {author} {\bibinfo {author} {\bibfnamefont {A.~D.}\ \bibnamefont
  {Hillier}}, \bibinfo {author} {\bibfnamefont {J.}~\bibnamefont
  {Quintanilla}}, \ and\ \bibinfo {author} {\bibfnamefont {R.}~\bibnamefont
  {Cywinski}},\ }\bibfield  {title} {\enquote {\bibinfo {title} {Evidence for
  time-reversal symmetry breaking in the noncentrosymmetric superconductor
  {LaNiC$_2$}},}\ }\href@noop {} {\bibfield  {journal} {\bibinfo  {journal}
  {Phys. Rev. Lett.}\ }\textbf {\bibinfo {volume} {102}},\ \bibinfo {pages}
  {117007} (\bibinfo {year} {2009})}\BibitemShut {NoStop}%
\bibitem [{\citenamefont {Biswas}\ \emph {et~al.}(2013)\citenamefont {Biswas},
  \citenamefont {Luetkens}, \citenamefont {Neupert}, \citenamefont
  {St{\"u}rzer}, \citenamefont {Baines}, \citenamefont {Pascua}, \citenamefont
  {Schnyder}, \citenamefont {Fischer}, \citenamefont {Goryo}, \citenamefont
  {Lees} \emph {et~al.}}]{BiswasTRS}%
  \BibitemOpen
  \bibfield  {author} {\bibinfo {author} {\bibfnamefont {P.}~\bibnamefont
  {Biswas}}, \bibinfo {author} {\bibfnamefont {H.}~\bibnamefont {Luetkens}},
  \bibinfo {author} {\bibfnamefont {T.}~\bibnamefont {Neupert}}, \bibinfo
  {author} {\bibfnamefont {T.}~\bibnamefont {St{\"u}rzer}}, \bibinfo {author}
  {\bibfnamefont {C.}~\bibnamefont {Baines}}, \bibinfo {author} {\bibfnamefont
  {G.}~\bibnamefont {Pascua}}, \bibinfo {author} {\bibfnamefont
  {A.}~\bibnamefont {Schnyder}}, \bibinfo {author} {\bibfnamefont
  {M.}~\bibnamefont {Fischer}}, \bibinfo {author} {\bibfnamefont
  {J.}~\bibnamefont {Goryo}}, \bibinfo {author} {\bibfnamefont
  {M.}~\bibnamefont {Lees}},  \emph {et~al.},\ }\bibfield  {title} {\enquote
  {\bibinfo {title} {{Evidence for superconductivity with broken time-reversal
  symmetry in locally noncentrosymmetric SrPtAs}},}\ }\href@noop {} {\bibfield
  {journal} {\bibinfo  {journal} {Phys. Rev. B}\ }\textbf {\bibinfo {volume}
  {87}},\ \bibinfo {pages} {180503} (\bibinfo {year} {2013})}\BibitemShut
  {NoStop}%
\bibitem [{\citenamefont {Brandt}(1988)}]{Brandt}%
  \BibitemOpen
  \bibfield  {author} {\bibinfo {author} {\bibfnamefont {E.~H.}\ \bibnamefont
  {Brandt}},\ }\bibfield  {title} {\enquote {\bibinfo {title} {{Flux
  distribution and penetration depth measured by muon spin rotation in
  high-${T}_{c}$ superconductors}},}\ }\href@noop {} {\bibfield  {journal}
  {\bibinfo  {journal} {Phys. Rev. B}\ }\textbf {\bibinfo {volume} {37}},\
  \bibinfo {pages} {2349--2352} (\bibinfo {year} {1988})}\BibitemShut {NoStop}%
\bibitem [{\citenamefont {Tinkham}(2004)}]{Tinkham}%
  \BibitemOpen
  \bibfield  {author} {\bibinfo {author} {\bibfnamefont {M.}~\bibnamefont
  {Tinkham}},\ }\href {https://books.google.com/books?id=VpUk3NfwDIkC} {\emph
  {\bibinfo {title} {Introduction to Superconductivity}}},\ Dover Books on
  Physics Series\ (\bibinfo  {publisher} {Dover Publications},\ \bibinfo {year}
  {2004})\BibitemShut {NoStop}%
\bibitem [{\citenamefont {Carrington}\ and\ \citenamefont
  {Manzano}(2003)}]{carrington}%
  \BibitemOpen
  \bibfield  {author} {\bibinfo {author} {\bibfnamefont {A.}~\bibnamefont
  {Carrington}}\ and\ \bibinfo {author} {\bibfnamefont {F.}~\bibnamefont
  {Manzano}},\ }\bibfield  {title} {\enquote {\bibinfo {title} {{Magnetic
  penetration depth of MgB$_2$}},}\ }\href@noop {} {\bibfield  {journal}
  {\bibinfo  {journal} {Physica C}\ }\textbf {\bibinfo {volume} {385}},\
  \bibinfo {pages} {205--214} (\bibinfo {year} {2003})}\BibitemShut {NoStop}%
\bibitem [{\citenamefont {Padamsee}, \citenamefont {Neighbor},\ and\
  \citenamefont {Shiffman}(1973)}]{padamsee}%
  \BibitemOpen
  \bibfield  {author} {\bibinfo {author} {\bibfnamefont {H.}~\bibnamefont
  {Padamsee}}, \bibinfo {author} {\bibfnamefont {J.}~\bibnamefont {Neighbor}},
  \ and\ \bibinfo {author} {\bibfnamefont {C.}~\bibnamefont {Shiffman}},\
  }\bibfield  {title} {\enquote {\bibinfo {title} {Quasiparticle phenomenology
  for thermodynamics of strong-coupling superconductors},}\ }\href@noop {}
  {\bibfield  {journal} {\bibinfo  {journal} {J. Low Temp. Phys.}\ }\textbf
  {\bibinfo {volume} {12}},\ \bibinfo {pages} {387--411} (\bibinfo {year}
  {1973})}\BibitemShut {NoStop}%
\bibitem [{\citenamefont {Niedermayer}\ \emph {et~al.}(2002)\citenamefont
  {Niedermayer}, \citenamefont {Bernhard}, \citenamefont {Holden},
  \citenamefont {Kremer},\ and\ \citenamefont {Ahn}}]{MgB2}%
  \BibitemOpen
  \bibfield  {author} {\bibinfo {author} {\bibfnamefont {C.}~\bibnamefont
  {Niedermayer}}, \bibinfo {author} {\bibfnamefont {C.}~\bibnamefont
  {Bernhard}}, \bibinfo {author} {\bibfnamefont {T.}~\bibnamefont {Holden}},
  \bibinfo {author} {\bibfnamefont {R.~K.}\ \bibnamefont {Kremer}}, \ and\
  \bibinfo {author} {\bibfnamefont {K.}~\bibnamefont {Ahn}},\ }\bibfield
  {title} {\enquote {\bibinfo {title} {{Muon spin relaxation study of the
  magnetic penetration depth in ${\mathrm{MgB}}_{2}$}},}\ }\href {\doibase
  10.1103/PhysRevB.65.094512} {\bibfield  {journal} {\bibinfo  {journal} {Phys.
  Rev. B}\ }\textbf {\bibinfo {volume} {65}},\ \bibinfo {pages} {094512}
  (\bibinfo {year} {2002})}\BibitemShut {NoStop}%
\bibitem [{\citenamefont {von Rohr}\ \emph {et~al.}(2019)\citenamefont {von
  Rohr}, \citenamefont {Orain}, \citenamefont {Khasanov}, \citenamefont
  {Witteveen}, \citenamefont {Shermadini}, \citenamefont {Nikitin},
  \citenamefont {Chang}, \citenamefont {Wieteska}, \citenamefont {Pasupathy},
  \citenamefont {Hasan} \emph {et~al.}}]{NbSe2}%
  \BibitemOpen
  \bibfield  {author} {\bibinfo {author} {\bibfnamefont {F.}~\bibnamefont {von
  Rohr}}, \bibinfo {author} {\bibfnamefont {J.-C.}\ \bibnamefont {Orain}},
  \bibinfo {author} {\bibfnamefont {R.}~\bibnamefont {Khasanov}}, \bibinfo
  {author} {\bibfnamefont {C.}~\bibnamefont {Witteveen}}, \bibinfo {author}
  {\bibfnamefont {Z.}~\bibnamefont {Shermadini}}, \bibinfo {author}
  {\bibfnamefont {A.}~\bibnamefont {Nikitin}}, \bibinfo {author} {\bibfnamefont
  {J.}~\bibnamefont {Chang}}, \bibinfo {author} {\bibfnamefont
  {A.}~\bibnamefont {Wieteska}}, \bibinfo {author} {\bibfnamefont
  {A.}~\bibnamefont {Pasupathy}}, \bibinfo {author} {\bibfnamefont
  {M.}~\bibnamefont {Hasan}},  \emph {et~al.},\ }\bibfield  {title} {\enquote
  {\bibinfo {title} {Unconventional scaling of the superfluid density with the
  critical temperature in transition metal dichalcogenides},}\ }\href@noop {}
  {\bibfield  {journal} {\bibinfo  {journal} {arXiv preprint arXiv:1903.05292}\
  } (\bibinfo {year} {2019})}\BibitemShut {NoStop}%
\bibitem [{\citenamefont {Guguchia}\ \emph {et~al.}(2015)\citenamefont
  {Guguchia}, \citenamefont {Amato}, \citenamefont {Kang}, \citenamefont
  {Luetkens}, \citenamefont {Biswas}, \citenamefont {Prando}, \citenamefont
  {von Rohr}, \citenamefont {Bukowski}, \citenamefont {Shengelaya},
  \citenamefont {Keller} \emph {et~al.}}]{GuguchiaNature}%
  \BibitemOpen
  \bibfield  {author} {\bibinfo {author} {\bibfnamefont {Z.}~\bibnamefont
  {Guguchia}}, \bibinfo {author} {\bibfnamefont {A.}~\bibnamefont {Amato}},
  \bibinfo {author} {\bibfnamefont {J.}~\bibnamefont {Kang}}, \bibinfo {author}
  {\bibfnamefont {H.}~\bibnamefont {Luetkens}}, \bibinfo {author}
  {\bibfnamefont {P.~K.}\ \bibnamefont {Biswas}}, \bibinfo {author}
  {\bibfnamefont {G.}~\bibnamefont {Prando}}, \bibinfo {author} {\bibfnamefont
  {F.}~\bibnamefont {von Rohr}}, \bibinfo {author} {\bibfnamefont
  {Z.}~\bibnamefont {Bukowski}}, \bibinfo {author} {\bibfnamefont
  {A.}~\bibnamefont {Shengelaya}}, \bibinfo {author} {\bibfnamefont
  {H.}~\bibnamefont {Keller}},  \emph {et~al.},\ }\bibfield  {title} {\enquote
  {\bibinfo {title} {{Direct evidence for a pressure-induced nodal
  superconducting gap in the Ba$_{0.65}$Rb$_{0.35}$Fe$_2$As$_2$
  superconductor}},}\ }\href@noop {} {\bibfield  {journal} {\bibinfo  {journal}
  {Nat. Commun.}\ }\textbf {\bibinfo {volume} {6}},\ \bibinfo {pages} {8863}
  (\bibinfo {year} {2015})}\BibitemShut {NoStop}%
\end{thebibliography}%

\end{document}